\documentclass[aps,prd,amsmath,amssymb,nofootinbib,showpacs]{revtex4}
\usepackage{color}
\usepackage{graphicx}
\usepackage{bm}
\usepackage{bbm}

\usepackage{amsfonts}
\usepackage{amsmath}
\usepackage{amssymb}
\usepackage{amsthm}

\usepackage{psfrag}
\usepackage{pstricks,pst-grad,color,shadow}
\usepackage{dsfont}
\usepackage{tikz}
\usetikzlibrary{shapes,arrows,backgrounds}
\usepackage{soul}
\usepackage{placeins}

\definecolor{darkgreen}{rgb}{0,0.73,0}

\newcommand{\komm}[2]{\ensuremath{\left[\,#1,#2\,\right]}}

\newcommand{\algebra}[1]{\mathfrak{#1}}
\newcommand{\group}[1]{\mathcal{#1}}
\newcommand{\gO}{\mathcal{O}}

\newcommand{\gU}{\mathcal{U}}

\newcommand{\Z}{\mathbb{Z}}

\newcommand{\vev}[1]{\left\langle #1 \right\rangle}

\newcommand{\trP}{\mathcal{P}}

\newcommand{\trnsp}{\mathsf{T}}

\newcommand{\ft}[2]{{\textstyle\frac{#1}{#2}}}
\DeclareMathOperator{\tr}{tr}

\renewcommand{\Re}{\operatorname{Re}}

\newcommand{\includeEPSTEX}[1]{\includegraphics{#1}}

\graphicspath{{./}{publPlots/}}
\begin{document}
\title{Phase diagram of the lattice $G_2$ Higgs Model}

\author{Bj\"orn H. Wellegehausen}
\author{Andreas Wipf}
\author{Christian Wozar}
\thanks{\texttt{bjoern.wellegehausen@uni-jena.de, wipf@tpi.uni-jena.de}
and \texttt{Christian.Wozar@uni-jena.de}}
\affiliation{Theoretisch-Physikalisches Institut,
Friedrich-Schiller-Universit{\"a}t Jena, Max-Wien-Platz 1, 07743
Jena, Germany}

\pacs{11.15.-q, 11.15.Ha, 12.38.Aw}

\begin{abstract}

\noindent
We study the phases and phase transition lines of 
the finite temperature $G_2$ Higgs model. Our work
is based on an efficient local hybrid Monte-Carlo algorithm
which allows for accurate measurements of expectation values,
histograms and susceptibilities.
On smaller lattices we calculate the phase diagram in
terms of the inverse gauge coupling $\beta$ and the 
hopping parameter $\kappa$.
For $\kappa\to 0$ the model reduces to $G_2$ gluodynamics 
and for $\kappa\to\infty$ to $SU(3)$ gluodynamics.
In both limits the system shows a first order confinement-deconfinement
transition. We show that the first order transitions at asymptotic
values of the hopping parameter are almost joined by a line of first
order transitions. A careful analysis reveals that there  exists a small 
gap in the line where the first order
transitions turn into continuous transitions or a cross-over region.
For $\beta\to\infty$ the gauge degrees of freedom are frozen
and one finds a nonlinear $O(7)$ sigma model
which exhibits a second order transition from a massive 
$O(7)$-symmetric to a massless $O(6)$-symmetric phase.
The corresponding second order line for large $\beta$ remains 
second order for intermediate $\beta$ until it comes close to the 
gap between the two first order lines. Besides this second order
line and the first order confinement-deconfinement transitions 
we find a line of monopole-driven bulk transitions which do not interfer 
with the confinement-deconfinment transitions.
\end{abstract}

\maketitle
\section{Introduction}
\noindent
Quarks and gluons are confined in mesons and baryons
and are not seen as asymptotic states of strong interaction. 
Understanding the dynamics of this confinement mechanism is one of the 
challenging problems in strongly coupled gauge theories.
Confinement is lost under extreme conditions:
when temperature reaches the QCD energy scale or the density
rises to the point where the average inter-quark separation
is less than $1$~fm, then hadrons are melted into their constituent
quarks. 

For gauge groups with a non-trivial center is the Polyakov loop 
\begin{equation}
P(\vec{x})=\tr \trP(\vec{x}),\quad \trP(\vec{x})=\frac{1}{N}\tr 
\left(\exp\; i\int_0^{\beta_T}
A_0(\tau,\vec{x}) \,d\tau\right),\quad\beta_T=\frac{1}{kT},\label{intro1}
\end{equation}
an order parameter for the  transition from the confined
to the unconfined phase in \emph{gluodynamics} (pure gauge theories).
Its thermal expectation value is related 
to the difference in free energy $F$ due to the presence
of an infinitely heavy test quark in the gluonic bath as 
\begin{equation}
\langle P\rangle \propto e^{-\beta F},
\end{equation}
such that $\langle P\rangle\neq 0$ in the unconfined high-temperature phase
and $\langle P\rangle= 0$ in the confined low-temperature phase.
Below the critical temperature is $\trP(\vec{x})$ uniformly
distributed over the group manifold and above the critical temperature 
it is in the neighborhood of a center-element. 
Near the transition point its dynamics is successfully described by effective
three dimensional scalar field models for the characters of $\trP(\vec{x})$
\cite{Svetitsky:1982gs,Yaffe:1982qf,Wozar:2006fi}. 
If one further projects the Polyakov loops onto the center of the 
gauge group, then one arrives at generalized Potts models 
describing the effective Polyakov-loop dynamics \cite{wipf:2006wj}. 

With matter in  the fundamental representation the center 
symmetry is \emph{explicitly broken} and  for all temperatures 
has $P$ a non-zero expectation value and points in the direction of a particular 
center element. Thus in the strict sense the Polyakov loop ceases to be
an order parameter for the center symmetry. On a microscopic scale
this is attributed to the breaking of the string connecting 
a static `quark anti-quark pair' when one tries to separate
the static charges \cite{Greensite:2003bk}. It breaks
via the spontaneous creation of dynamical quark anti-quark pairs which in turn
screen the individual static charges.

To clarify the relevance of the center symmetry  for confinement it suggests itself 
to study  gauge theories for which the gauge group has a trivial center.
Then the Polyakov loop ceases to be an order parameter even 
in the absence of dynamical matter since the strings connecting external 
charges can break via the spontaneous creation of dynamical `gluons'. 
The smallest simple and simply connected Lie group with a trivial center is 
the $14$ dimensional exceptional Lie group $G_2$. This is one reason 
why $G_2$ gauge theory with and without Higgsfields has been investigated 
in series of papers \cite{Holland:2003kg, Holland:2003jy,Pepe:2006er,Greensite:2006sm,Danzer:2008bk,
Maas:2007af}. 
Although there is no symmetry reason for a deconfinement phase transition
in  $G_2$ gluodynamics it has been conjectured that a first order 
deconfinement transition without order parameter exists.
In this context confinement refers to confinement at intermediate scales, where 
a Casimir scaling of string tensions has been detected in \cite{Liptak:2008gx}. 
Although the threshold energy for string breaking in $G_2$ gauge theory 
is rather high, string breaking has been seen in $3$ dimensional
$G_2$ gluodynamics in \cite{Wellegehausen:2010ai}.

\begin{figure}[htb]
\scalebox{1}{\includeEPSTEX{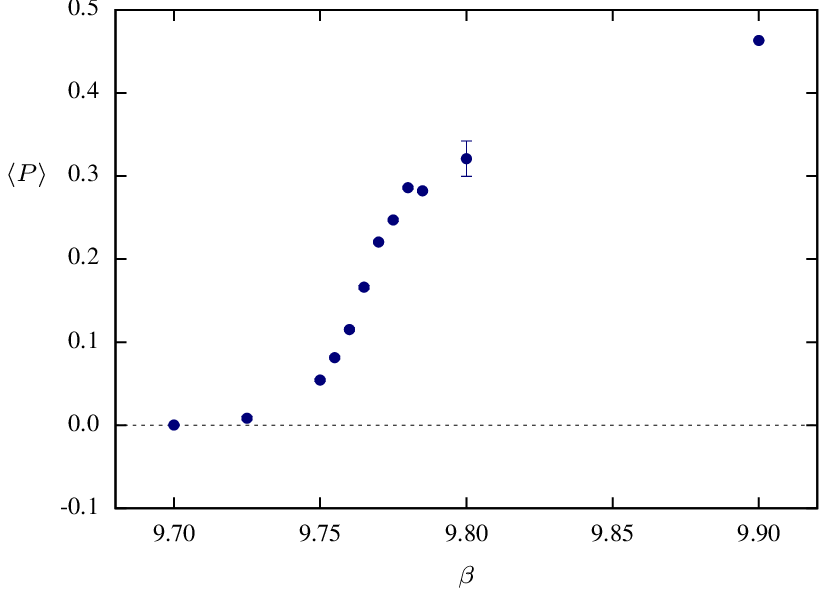}}\hskip10mm
\scalebox{1}{\includeEPSTEX{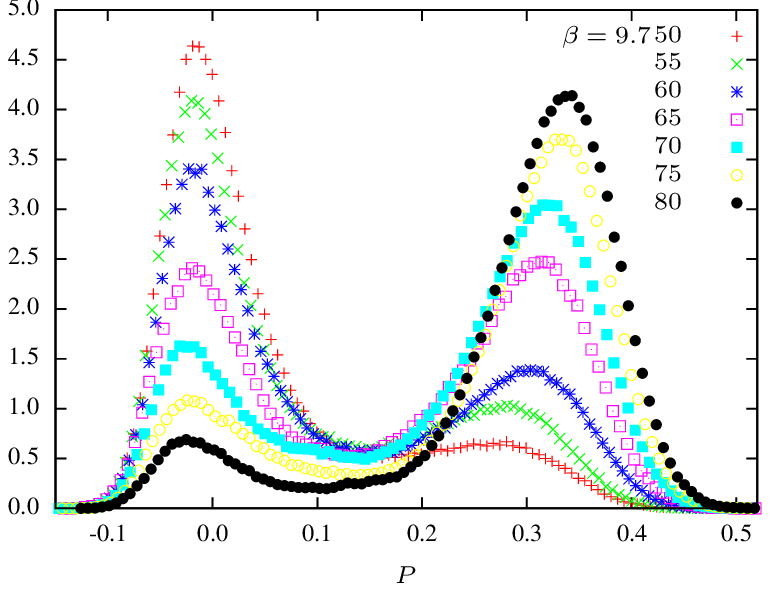}}
\caption{Phase transition on a $16^3 \times 6$ lattice in terms of the Polyakov
loop in the fundamental representation of $G_2$. The rapid
change of the Polyakov loop with $\beta=1/g^2$ (left panel)
and histograms of the Polyakov loop for $\beta$ in
the vicinity of $\beta_c$ (right panel) point to a first order
transition.}
\label{fig:ymTransition}
\end{figure}

The gauge group $SU(3)$ of strong interaction is a subgroup of $G_2$ and this
observation has interesting consequences, as pointed out
in \cite{Holland:2003jy}. With a Higgs field in the fundamental
$7$ dimensional representation one can break the $G_2$ gauge symmetry 
to the $SU(3)$ symmetry via the Higgs mechanism. 
When the Higgs field in the action
\begin{equation}
S[A,\phi]=\int d^4x\left(\frac{1}{4g^2}\tr F_{\mu\nu}F^{\mu\nu} 
+\frac{1}{2}(D_\mu\phi,D_\mu\phi) +V(\phi)\right)\label{contaction}
\end{equation}
picks up a vacuum expectation value $v$, then $6$ gauge bosons acquire 
a mass proportional to $v$ while the $8$ gluons belonging to $SU(3)$ remain 
massless. The massive gauge bosons are removed from the spectrum 
for $v\to\infty$. In this limit $G_2$ Higgs model reduces to $SU(3)$ Yang-Mills theory.
Even more interesting, for intermediate and large values of $v$
the $G_2$ Yang-Mills-Higgs (YMH) theory mimics $SU(3)$ gauge theory with 
dynamical `scalar quarks'. The masses  of these `quarks' and the length scale
at which string breaking occurs increase with increasing $v$.
The Polyakov loop serves as \emph{approximate order parameter} 
separating the confined from the unconfined phases with a rapid change 
at the transition or crossover. This rapid change is depicted in
Fig.~\ref{fig:ymTransition} which shows the expectation value of $P$ for $G_2$
gluodynamics as function of the inverse gauge coupling $\beta=1/g^2$.

In an earlier work we derived a $3$ dimensional \emph{effective theory}
for the dynamics of the Polyakov loop for finite temperature $G_2$ gluodynamics
and analyzed the resulting Landau-type theory with the help of elaborate
Monte Carlo simulations \cite{Wellegehausen:2009rq}. Already the leading order 
effective Polyakov loop model exhibits a rich phase structure with symmetric, 
ferromagnetic, and anti-ferromagnetic phases.

\begin{figure}[h]
\scalebox{0.6}{\includeEPSTEX{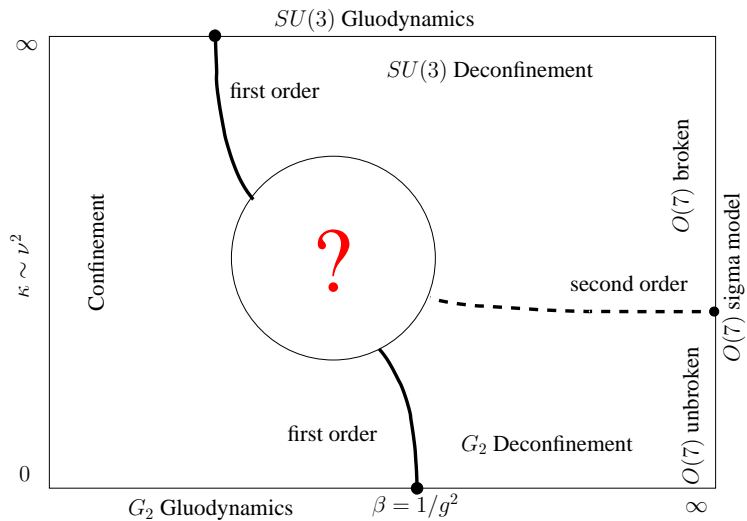}}
\caption{Expected phase diagram in the parameter space $(1/g^2,\kappa)$
(taken from \cite{Pepe:2006er}).}
\label{fig:sketchphases}
\end{figure}

In the present paper we investigate the phase structure of microscopic $G_2$ YMH 
lattice theory with a Higgs field in the  $7$ dimensional representation. 
The corresponding lattice action for the $G_2$ valued link variables and a
normalized Higgs field with $7$ real components reads
\begin{equation}
S_{\rm YMH}[\,\gU,\Phi] = \beta \sum \limits_\square \left( 1-\frac{1}{7} \tr \Re
\gU_\square \right)-\kappa \sum \limits_{x,\mu} \Phi_{x+\hat\mu}\, \gU_{x,\mu}
\Phi_{x},\quad \Phi_x\cdot\Phi_x=1,\label{latticeaction}
\end{equation}
and depends on the inverse gauge coupling $\beta$ and the hopping parameter
$\kappa$. For $\beta\to\infty$ the gauge bosons 
decouple and the theory reduces to an $O(7)$ invariant nonlinear sigma model
which is expected the have a  second order (mean field) symmetry 
breaking transition down to $O(6)$. The mean field prediction for 
the critical coupling is $\kappa_{c,\rm mf}=7/8$ and this value
bounds $\kappa_c$ from below \cite{simon80}. In the limit $\kappa=0$ we recover
$G_2$ gluodynamics with a first order deconfinement phase
transition, in agreement with the findings in \cite{Cossu:2008}. 
In the other extreme case $\kappa\to\infty$ we end up with $SU(3)$ gluodynamics
with a weak first order deconfinement transition. The known transitions
in the limiting cases $\kappa\to 0,\;\kappa\to\infty$ or $\beta\to\infty$
are depicted in Fig.~\ref{fig:sketchphases}.
If $\kappa$ is lowered from $\infty$ then in addition to the $8$ gluons 
of $SU(3)$, the $6$ additional gauge bosons
of $G_2$ with decreasing mass begin to participate in the dynamics.
Similarly as dynamical quarks and anti-quarks, they transform in
the representations $\{3\}$ and $\{\bar 3\}$ of $SU(3)$ and thus explicitly
break the $\Z_3$ center symmetry. As in \emph{QCD} they are expected to
weaken the deconfinement phase transition. Thus it has been conjectured
in \cite{Pepe:2006er} that there may exist a critical endpoint where the 
transition disappears. 

In the following section we shall briefly recall those facts about $G_2$
representations which are relevant for the present work. In Sec.~\ref{sec:algo}
some algorithmic aspects are reviewed. A more detailed presentation can be found
in our earlier paper \cite{Wellegehausen:2010ai}. Sec.~\ref{sec:phaseDiagram} 
contains our Monte-Carlo results for the phase diagram in the 
$(\beta,\kappa)$ plane. 
We find that the two first order lines emanating from the deconfinement
transitons in $G_2$ and $SU(3)$ gluodynamics at $\kappa=0$ 
and $\kappa=\infty$ end in the vicinity of $(\beta,\kappa)=(9.4,1.6)$ on 
a $6\times 16^3$ lattice. Sec.~\ref{sec:transitionLinesNear} contains the results 
of our high statistics simulations for histograms and susceptibilities
in the small region in parameter space where the two first order
lines are either connected by a second order line or leave open
a gap which smoothly connects the confinend and deconfined
phases. Our data are consistent with the conjectured critical 
endpoints attached to the two first order  lines.
For large $\beta$ a second order transition line which 
separates the $O(7)$ and $O(6)$ sigma models comes close to
the first order deconfinement transition lines.
The phases and transition lines are localized and analysed with high 
statistics simulations of the Polyakov loop distribution and susceptibility,
plaquette and Higgs action susceptibilities, and finally with derivatives
of the mean action with respect to the hopping parameter.
Besides the transition lines indicated in Fig.~\ref{fig:sketchphases}
there exists another line of monopole driven bulk transitions.
This line emanates from the bulk crossover in pure $G_2$-gluodynamics 
at $\beta=9.45$  \cite{Cossu:2008}.

\section{The group $\bm{G_2}$}
\label{sect:g2}
\noindent
The exceptional Lie group $G_2$ is the smallest Lie group in
the Cartan classification which is simply connected and has
a trivial center.  The two fundamental representations
are the $7$ dimensional defining representation $\{7\}$ and 
the $14$ dimensional adjoint representation $\{14\}$.
One may view the elements of the  representation $\{7\}$ as matrices 
in the defining representation of $SO(7)$, subject to seven independent cubic 
constraints, see \cite{Holland:2003jy}. For example, 
the defining representation $\{7\}$  of $SO(7)$ turns into an irreducible representation
of $G_2$, whereas the adjoint representation $\{21\}$ of $SO(7)$ branches into the 
two fundamental representation $\{14\}$ and $\{7\}$ of $G_2$. 
The gauge group of strong interaction is a subgroup of $G_2$ 
and the corresponding coset space is a sphere \cite{Macfarlane:2002hr},
\begin{equation}
G_2/SU(3) \sim S^6.
\end{equation}
This means that every element $\gU$ of $G_2$ can be written as
\begin{equation}
\gU=\group{S} \cdot \group{V} \quad \text{with} \quad\group{S} \in G_2/SU(3)
\quad \text{and} \quad \group{V} \in SU(3),\label{decomposition}
\end{equation}
and we shall use this decomposition to speed up our numerical simulations.


Quarks in $G_2$ transform under the $7$ dimensional fundamental representation,
gluons under the $14$ dimensional fundamental (and adjoint) representation.
To better understand $G_2$ gluodynamics we recall the decomposition of
tensor products
\begin{equation}
\label{eq:representationsG2}
\begin{aligned}
\{7\} \otimes \{7\}&=\{1\} \oplus \{7\} \oplus \{14\} \oplus \{27\}, \\ 
\{7\} \otimes \{7\} \otimes \{7\}&=\{1\} \oplus 4 \cdot \{7\} \oplus 2 \cdot \{14\} \oplus 3 \cdot \{27\} 
\oplus 2 \cdot \{64\} \oplus \{77'\}, \\
\{14\} \otimes \{14\}&=\{1\} \oplus \{14\} \oplus \{27\} \oplus \{77\} \oplus
\{77'\}, \\
\{14\} \otimes \{14\}\otimes \{14\}&=\{1\}\oplus \{7\} \oplus 5\cdot \{14\}\oplus 
3\cdot\{27\}\oplus\dotsb.
\end{aligned}
\end{equation}
These decompositions show similarlies to QCD:
two quarks, three quarks, two gluons and three gluons can build
colour singlets -- mesons, baryons and glueballs.
In $G_2$ gauge theory three gluons can screen the colour charge of a single quark,
\begin{equation}
\{7\} \otimes \{14\} \otimes \{14\} \otimes \{14\}=\{1\} \oplus \dotsb,
\end{equation}
and this explains why the string between two external charges  in the $\{7\}$
representation will break for large charge separations.  The two remnants are
colour blind glue lumps. The same happens for two external charges in the 
adjoint representation. In a previous work we did observe string breaking at the expected 
separation between the two charges \cite{Wellegehausen:2010ai}.

The  $G_2$ gauge symmetry can be broken to $SU(3)$ with the help of a
Higgs field in the  $7$ dimensional representation. For $\kappa\to\infty$ the
factor $\group{S}$ in the decomposition \eqref{decomposition} is frozen and we
end up with an $SU(3)$ gauge theory with rescaled gauge coupling for 
the factor $\gU$. With respect to the unbroken subgroup $SU(3)$ 
the fundamental representations $\{7\}$ and $\{14\}$ branch into the 
following irreducible $SU(3)$ representations:
\begin{equation}
\begin{aligned}
\{7\}&\longrightarrow\{3\}\oplus \{\bar{3}\} \oplus \{1\}, \\
\{14\}&\longrightarrow\{8\} \oplus \{3\} \oplus \{\bar{3}\}.
\end{aligned}
\end{equation}
The Higgs field branches into a scalar quark, scalar anti-quark and singlet with 
respect to $SU(3)$. Similarly, the $G_2$ gluons branch into massless
$SU(3)$ gluons and additional gauge bosons with respect to $SU(3)$. The latter
eat up the non-singlet scalar fields such that the spectrum in the broken phase
consists of $8$ massless gluons, $6$ massive gauge bosons and one
massive Higgs particle.

\section{Algorithmic considerations}
\label{sec:algo}
\subsection{Equations of motion for local hybrid Monte-Carlo}
\noindent
In this work we employ a local version of the hybrid Monte-Carlo (HMC) algorithm
where single site and link variables are evolved in a HMC style \cite{Marenzoni:1993im}. 
The algorithm assumes a local interaction and hence applies to all purely
bosonic theories. The implementation for the $G_2$ Higgs model is a 
mild generalization of the algorithm used in our previous work on $G_2$ 
gluodynamics \cite{Wellegehausen:2010ai}. We use a local hybrid Monte-Carlo
(LHMC) algorithm for several good reasons: First  there is no low
Metropolis acceptance rate even for large hopping parameters. 
More precisely, in a heat bath algorithm combined with an 
over-relaxation we would need two Metropolis steps in each
update for $\kappa> 0$ which for large $\kappa$ may lead to low 
acceptance rates. With the LHMC-algorithm we can avoid this problem
and deal with arbitrary values of $\kappa$. Autocorrelation times can 
be controlled (in certain ranges) by the integration time in the molecular 
dynamics part of the HMC algorithm.
Second, the formulation is given entirely in terms of Lie group and Lie 
algebra elements and there is no need to back-project onto the group. 
For $G_2$ it is possible to use a real representation and in addition 
an analytical expression for the involved exponential maps 
from the algebra  to the group.  These maps allow for a fast implementation 
of the LHMC algorithm.

 This algorithm has been essential for obtaining the accurate results 
in the present work. Since we developed and used
the first implementation for $G_2$ it may be useful to sketch how it works for
this exceptional group. More details can be found in \cite{Wellegehausen:2010ai}.
For $G_2$ YMH lattice theory the (L)HMC algorithm is based on a fictitious
dynamics for the link-variables on the $G_2$ manifold and the normalized Higgs
field on the $6$-sphere. The ``free evolution'' on a semisimple group is the
Riemannian geodesic motion with respect to the Cartan-Killing metric
\begin{equation}
ds^2_G=\kappa\tr\left(d\gU \gU^{-1}\otimes d\gU\gU^{-1}\right).
\end{equation}
In a (L)HMC dynamics the interaction term is given by the YMH
action \eqref{latticeaction} of the underlying lattice gauge theory and hence
it is natural to derive the HMC dynamics from a Lagrangian of the form
\begin{equation}
L_{\rm HMC}=-\frac{1}{2}  \sum \limits_{x,\mu}\tr\left(\dot{\gU}_{x,\mu}\gU_{x,\mu}^{-1}\right)^2
+K(\Phi, \dot\Phi)
-S_\text{YMH}[\,\gU,\Phi],
\end{equation}
where `dot' denotes the derivative with respect to the fictitious time 
parameter $\tau$ and $K(\Phi,\dot\Phi)$ is a kinetic term for the Higgs field.
%
To update the normalized Higgs field we set
\begin{equation}
 \Phi_x=\gO_x\Phi_0\quad\hbox{with}\quad \gO_x\in SO(7)
\end{equation}
and constant $\Phi_0$. The change of variables $\Phi_x\to \gO_x$
converts the  induced measure on $S^6\subset R^7$ into the Haar measure of $SO(7)$.
Without interaction the rotation matrices $\gO_x$ will evolve freely on the group
manifold $SO(7)$ such that in terms of the $(\gU,\gO)$ variables we choose as
Lagrangian for the HMC dynamics
\begin{equation}
L=-\frac{1}{2}  \sum \limits_{x,\mu}\tr\left(\dot{\gU}_{x,\mu}\gU_{x,\mu}^{-1}\right)^2
-\frac{1}{2}\sum\limits_x  \tr\left(\dot\gO_x\gO^{-1}_x\right)^2
-S_{\rm YMH}[\,\gU,\gO]\,.
%
\end{equation}
The Lie algebra valued  fictitious momenta conjugated to the link variable $\gU_{x,\mu}$ and
site variable $\gO_x$ are given by
\begin{equation}
\algebra{P}_{x,\mu}=\frac{\partial L}{\partial
\big(\dot{\gU}_{x,\mu}\gU_{x,\mu}^{-1}\big)}=-
\dot{\gU}_{x,\mu}\gU_{x,\mu}^{-1}\quad,\quad
\algebra{Q}_x= \frac{\partial L}{\partial
\big(\dot{\gO}_{x}\gO_{x}^{-1}\big)}=-\dot\gO_x\gO_x^{-1}.
\label{hmcequations1}
\end{equation}
The Legendre transform yields the following pseudo-Hamiltonian
\begin{equation}
H=-\frac{1}{2} \sum_{x,\mu}\tr \algebra{P}_{x,\mu}^2
-\frac{1}{2}\sum_x \tr \algebra{Q}_x^2
+S_\text{YMH}[\,\gU,\gO].
%
\end{equation}
Note that for real $\gU_{x,\mu}$ and $\gO_x$ the momenta are antisymmetric
such that both kinetic terms are positive. The equations of motion for the momenta 
are obtained by varying the Hamiltonian. The variation of $S_\text{\rm
YMH}[\,\gU,\gO]$ with respect to a fixed link variable $\gU_{x,\mu}$ yields 
the staple variable $R_{x,\mu}$, the sum of triple products of elementary link variables closing 
to a plaquette with the chosen link variable. Setting
\begin{equation}
\delta\algebra{P}_{x,\mu}=\dot{\algebra{P}}_{x,\mu}d\tau,\quad
\delta \gU_{x,\mu}=\dot\gU_{x,\mu}d\tau=-\algebra{P}_{x,\mu}\gU_{x,\mu}d\tau
\end{equation}
with similar expressions for the momentum and field variables
$\algebra{Q}_x$ and $\gO_{x}$ in the Higgs sector yields
for the variation of the HMC Hamiltonian
\begin{equation}
\label{HMChamiltonian}
\delta H 
= -\sum \limits_{x,\mu} \tr \algebra{P}_{x,\mu}\big\{\dot{\algebra{P}}_{x,\mu}- F_{x,\mu}\big\}
-\sum_x \tr \algebra{Q}_x \big\{\dot{\algebra{Q}}_x- G_x\big\}
\end{equation}
with the following ``forces'' in the gauge  and Higgs sector
\begin{equation}
\label{HMCForce}
F_{x,\mu}= \frac{\beta}{14}
\left(\gU_{x,\mu}R_{x,\mu}-R_{x,\mu}^\dagger \gU^\dagger_{x,\mu}\right)
+\kappa (\gU_{x,\mu}\phi_x)\phi^\trnsp_{x+\mu}
,\quad
G_x=\kappa\phi_x\Big(\sum\nolimits_{y:x} \gU_{xy}\,\phi_y\Big)^\trnsp,
\end{equation}
where the last sum extends over all nearest neighbors $y$ of $x$
and $U_{xy}$ denotes the parallel transporter from $y$ to $x$.
The variational principle implies that the projection of the terms between
curly brackets onto the Lie algebras $\algebra{g}_2$ and $\algebra{so}(7)$ vanish,
\begin{equation}
\dot{\algebra{P}}_{x,\mu}=
F_{\mu,x}\big\vert_{\algebra{g}_2}\quad,\quad
\dot{\algebra{Q}}_x=G_x\big\vert_{\algebra{so}(7)}.\label{hmcequations2}
\end{equation}
The equations \eqref{hmcequations1} and \eqref{hmcequations2} determine
the fictitious dynamics of the lattice fields in the (L)HMC algorithm.
Choosing a trace-orthonormal basis $\{T_a\}$ of $\algebra{g}_2$
the LHMC equations in the gauge sector read
\begin{equation} \dot{\gU}_{x,\mu}=-
\algebra{P}_{x,\mu}\gU_{x,\mu}\quad\hbox{and}\quad
\dot{\algebra{P}}_{x,\mu}=\sum \limits_a \tr \left(F_{x,\mu}T_a\right)T_a
\end{equation}
with force $F_{x,\mu}$ defined in \eqref{HMCForce}.
In the Higgs sector they take the form
\begin{equation}
 \dot{\gO}_x=-
\algebra{Q}_{x}\gO_x\quad \text{and} \quad
\dot{\algebra{Q}}_{x}=\sum \limits_b \tr \left(G_{x}\tilde T_b\right)\tilde T_b
\end{equation}
with trace-orthonormal basis $\{\tilde T_b\}$ of $\algebra{so}(7)$ and force
$G_{x}$ defined in \eqref{HMCForce}.
\subsection{Numerical solutions of YMH-dynamics}
We employ a time reversible leap frog integrator which uses the integration
scheme
\begin{equation}
 \begin{aligned}
\algebra{P}_{x,\mu}(\tau+\ft12\delta\tau)&=\algebra{P}_{x,\mu}(\tau)
+\ft12\delta\tau\, \dot{\algebra{P}}_{x,\mu}(\tau)\\
\gU_{x,\mu}(\tau+\delta\tau)&=\exp\left\{-\delta\tau\, \algebra{P}_{x,\mu}(\tau+\ft12\delta \tau)\right\}\gU_{x,\mu}(\tau)\\
\algebra{P}_{x,\mu}(\tau+\delta\tau)&=\algebra{P}_{x,\mu}(\tau+\ft12\delta\tau)+\ft12\delta\tau\,
\dot{\algebra{P}}_{x,\mu}(\tau+\delta\tau),
\end{aligned}
\end{equation}
and similarly for the variables $(\gO_x,\algebra{Q}_x)$ in the Higgs sector.
The `time' derivative of $\algebra{P}(\tau+\delta\tau)$ in the last step is
given in terms of the already known group valued field at $\tau+\delta \tau$
via the equations of motion.
Clearly, to calculate $\gU$ and $\gO$ at time $\tau+\delta \tau$ a fast implementation
of exponential maps is required. In the Higgs sector the map $\algebra{so}(7)\to
SO(7)$ is computed via the Cayley-Hamilton theorem. For small values of the hopping parameter $\kappa$ 
the step size and integration length for the integration may be chosen as in
the gauge field integrator.  For an efficient and fast computation of the exponential map 
$\algebra{g}_2\to G_2$ we exploit the \emph{real} embedding $\group{V}$ of the
representation $3\oplus\bar3$ of $SU(3)$ into $G_2$,
\begin{equation}
\gU=\group{S} \cdot \group{V}(\group{W}) \quad \text{with} 
\quad \group{S} \in G_2/SU(3),\quad \group{W}\in SU(3).\label{groupdecomp}
\end{equation}
For a given time step $\delta \tau$ the factorization will be
expressed in terms of the Lie algebra elements with the help of  the exponential maps,
\begin{equation}
\exp \left \lbrace \delta \tau\, \algebra{u} \right \rbrace=\exp \left \lbrace
\delta \tau\, \algebra{s} \right \rbrace\cdot \exp \left \lbrace \delta \tau\,
\algebra{v} \right \rbrace \quad \text{with generators} \quad \algebra{u}\in\algebra{g}_2,\;\;
\algebra{v} \in \group{V}_*(\algebra{su}(3)).\label{expalg}
\end{equation}
The exponential maps for the two factors can be calculated efficiently,
see \cite{Wellegehausen:2010ai}. But in the numerical integration 
we need the exponential map for elements $\algebra{u}\in\algebra{g}_2$. 
These elements are related to the generators $\algebra{s}$ and
$\algebra{v}$ used in the factorization by the 
Baker-Campbell-Hausdorff formula,
\begin{equation}
\delta \tau \, \algebra{u}=\delta \tau
\left(\algebra{s}+\algebra{v}\right)+\frac{1}{2}\delta \tau^2
\komm{\algebra{s}}{\algebra{v}}+ \cdots.\label{baker}
\end{equation}
For a second order integrator the approximation
(\ref{baker}) may be used in the exponentiations needed to calculate $\group{V}$
and $\group{S}$. This approximation leads to a violation of energy conservation 
which is of the same order as the violation one finds with a second order integrator. 
To sum up, a LHMC sweep consists of the following steps: 
\begin{enumerate}
\item Gaussian draw for the momentum variables on a given site and link,
\item  Integration of the equations of motion for the given site and link,
\item Metropolis accept/reject step,
\item Repeat these steps for all sites and links of the lattice.
\end{enumerate}
 This local version of the HMC does 
not suffer from an extensive $\delta H\propto V$ problem such that already
a second order symplectic (leap frog) integrator allows
for sufficiently large time steps $\delta \tau$. For a large range of couplings
$(\beta,\kappa)$ in our simulations an integration length of $T=0.75$ with a step size of $\delta \tau=0.25$ is optimal for minimal autocorrelation times and a small number of thermalisation sweeps. Acceptance
rates of more than $99\%$ are reached. To compare
the performances of our LHMC algorithm with the usually 
used heat-bath algorithm we estimated the computation time of 
the different parts in the LHMC-algorithm in units given by 
the average computation time for one staple in $\Delta S_{\gU}$. 
On an Intel Corei7 CPU the latter is approximately $4\,\mu\text{s}$ 
for a $12^3 \times 6$ lattice.

In Table \ref{tab:performance1} we listed the times needed
to change the gauge or Higgs action during a single
update of one link or one Higgs field variable, the
time for both integrators without exponential map and 
separately the computation time for a single exponential map.
Most time is spent with calculating the exponential maps for
$SO(7)$.
\begin{table}[htb]
\begin{tabular}{p{2cm}p{1.5cm}p{1.5cm}p{1.5cm}p{1.5cm}p{1.5cm}p{1.6cm}}
Part& $\Delta S_{\gU}$  &$\Delta S_{\gO}$   &integr. $\gU$  &integr. $\gO$ &
exp($G_2$) & exp($SO_7$) \\ \hline
pure gauge & $1.00$ & - & $1.34$ & - & $0.42$ & -\\ 
gauge Higgs & $1.03$ & $0.43$ & $1.74$ & $1.00$ &
$0.40$ & $4.97$
\end{tabular}
\caption{Computation times normalized to $\Delta S_{\gU}$ (staple).}
\label{tab:performance1}
\end{table}
Note that during the calculation of one exponential map for $SO(7)$ 
the CPU calculates about $10$ exponential maps for $G_2$. Table
\ref{tab:performance2} compares the total time-contributions to one 
configuration with those of the heat-bath algorithm with overrelaxation. 
We see that for pure gauge theories the standard heat-bath algorithm 
with overrelaxation is only two times faster as the LHMC algorithm.
\begin{table}[htb]
\begin{tabular}{p{2cm}p{1.2cm}p{1.2cm}p{1.2cm}p{1.2cm}p{1.2cm}p{1.6cm}
p{3.2cm}p{1.8cm}}
Part& $\Delta S_{\gU}$  &$\Delta S_{\gO}$&integr. $\gU$  &integr. $\gO$ &
exp($G_2$) & exp($SO_7$) & $\text{total time}/V\cdot d\cdot \text{Config.}$ &
heat-bath \\ \hline 
pure gauge & $1.00$ & - & $1.34$ & - & $1.26$ & - & $3.60$
& $\approx 2$\\ 
gauge Higgs & $1.03$ & $0.11$ & $1.74$ & $0.25$ & $1.20$ &
$3.72$ & $8.05$ & -
\end{tabular}
\caption{Total time contribution to one LHMC configuration compared to heat-bath
algorithm.}
\label{tab:performance2}
\end{table}

\section{The phase diagram of the $G_2$ Higgs model: overview}
\label{sec:phaseDiagram}
With the help of the local HMC algorithm sketched previously
we calculated several relevant observables to probe the
phases and phase transition lines in the $(\beta,\kappa)$ plane.
First we present the phase diagram obtained on small lattices.
For vanishing $\kappa$ we are dealing with $G_2$ gluodynamics
which shows a first order finite temperature deconfinement phase transition.
The transition is discontinuous since there is a large mismatch
of degrees of freedom in the confined and unconfined phases.
At the other extreme value $\kappa=\infty$ six of the fourteen gauge bosons decouple 
from the dynamics and we are left with  $SU(3)$ gluodynamics, 
which shows a first order deconfinement phase transition as well. 
The question arises whether the first order
transitions in $G_2$ and $SU(3)$ gluodynamics  are connected 
by a unbroken line of first order transitions or whether there 
are two critical endpoints. In the latter case the confined and 
unconfined phases could be connected continuously. 
On the other hand, for arbitrary $\kappa$ but $\beta\to\infty$ the 
gauge degrees of freedom decouple from the dynamics and one is left with a 
nonlinear $O(7)$-sigma model. We expect that the $O(7)$-symmetry is 
spontaneously broken to $O(6)$ for sufficiently large values of the hopping 
parameter and that this transition is of second order.

In order to localize the confinement-deconfinement transition line(s) we first
measured the Polyakov loop expectation value as (approximate)
order parameter for confinement on a small $12^3 \times 2$-lattice
in a large region of parameter space ($\beta=5 \dots 10$, $\kappa=0 \dots 10^4$).
For $\kappa\gg 1$  the Polyakov loop takes its values in the reducible 
representation $\{3\}\oplus\{\bar 3\}\oplus\{1\}$  of $SU(3)$ and
\begin{equation}
\langle P\rangle \approx 1+\langle P+\bar P\rangle_{SU(3)}.
\end{equation}
Thus, for large $\kappa$ we should find $\vev{P}\approx 1$ in the confining phase
and $\vev{P}\approx 7$ or $ \vev{P}\approx -2$
in the unconfined phase where $P$ is near one of the three center-element of
$SU(3)$.
We eliminate the ambiguity of assigning a value to the Polyakov loop in the 
unconfined phase by mapping values with $\vev{P}<1$ to  $3-2\vev{P}$.

\begin{figure}[htb]
\scalebox{1.6}{\includeEPSTEX{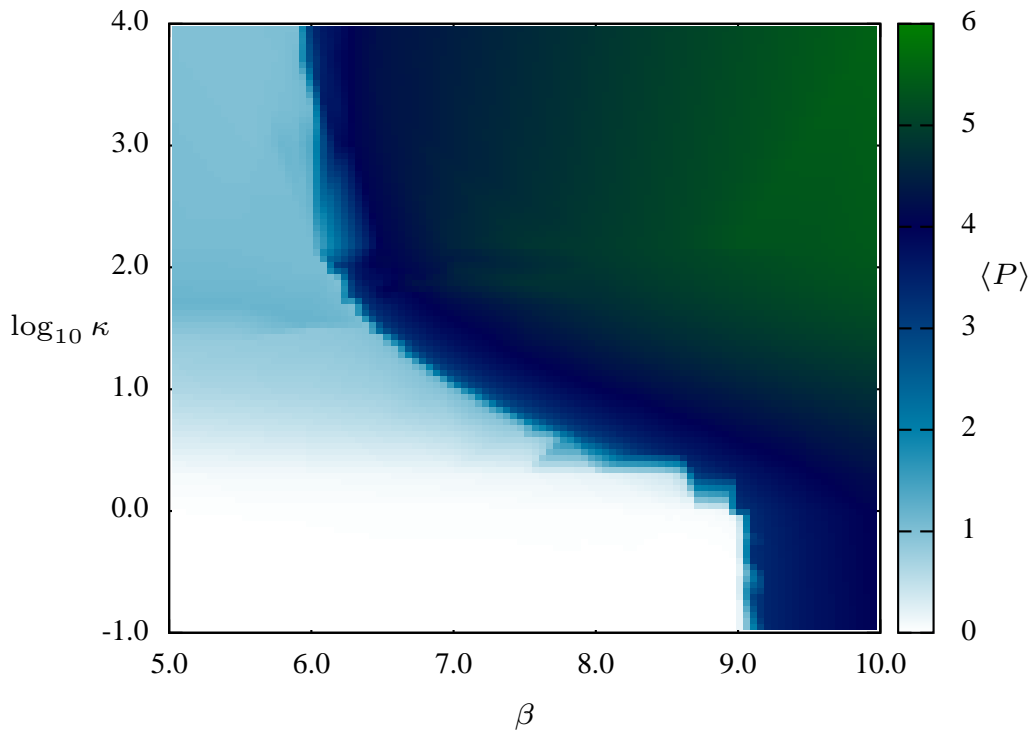}}
\caption{Expectation values of $P$  in the coupling constant plane and on a small $12^3 \times 2$ lattice}
\label{fig:phasediagram12}
\end{figure}

The result for $\vev{P}$ is depicted in Fig.~\ref{fig:phasediagram12}.
We see that in the confining phase the expectation value varies from $0$ to $1$ 
when the hopping parameter increases. For large values of $\beta$ in the unconfined
phase the Polyakov loop is near the identity or (for large $\kappa$) near one
of the three center-elements of $SU(3)$.
On the small lattice the Polyakov loop jumps along a continuous
curve connecting the confinement-deconfinement transitions of pure $G_2$ and 
pure $SU(3)$ gluodynamics. This suggests that there exists a connected first order 
transition curve all the way from $\kappa=0$ to $\kappa=\infty$.
To see whether this is indeed the case we performed high-precision
simulations on larger lattices. A careful analysis of histograms and 
susceptibilities for Polyakov loops and the Higgs action shows that
the first order lines beginning at $\kappa=0$ and at $\kappa=\infty$
do not meet. This happens in a rather small region in parameter 
space such that the two first order lines almost meet.
They may be connected by a line of continuous transitions
or in-between there may exists a window connecting the confined
and unconfined phases smoothly.

For $\beta\to\infty$ we are left with a nonlinear $O(7)$ sigma model with action
\begin{align}
S_\sigma = -\kappa \sum \limits_{x,\mu} \Phi_{x+\hat\mu}\Phi_{x}\,,\label{sigmaaction}
\end{align}
and this model shows a second order transition at a critical coupling
$\kappa_c$ from a $O(7)$ symmetric to a $O(6)$ symmetric phase. 
To see how this transition continues to finite values of $\beta$ we measured
the expectation values $\langle\gO_P\rangle$ and 
$\langle\gO_H\rangle$ of the (averaged) plaquette variable and Higgs action
\begin{align}
\gO_P=\frac{1}{7 \cdot 6 \cdot V}\sum_\square\Re \tr
\gU_\square\quad\hbox{and}\quad \gO_H=\frac{1}{V} \sum \limits_{x\mu} \Phi_{x+\hat\mu}\, \gU_{x,\mu}
\Phi_{x}\,.\label{operators}
\end{align}
and the corresponding  susceptibilities
\begin{align}
\chi(\gO)=V\left(\langle \gO^2\rangle-\langle\gO\rangle^2\right)\,.
\end{align}
The finite size scaling theory predicts that near the transition point the maximum of the susceptibilities scales with the volume to the power of the corresponding 
critical exponent $\gamma$
\begin{align}
\chi(\gO)\sim a L^{\gamma/\nu}+b\, ,
\end{align}
where $\nu$ is the critical exponent related to the divergence
of the correlation length. For a first order phase transition we expect the
susceptibility peak to scale linearly with the spatial volume (since $N_t$ is fixed).
More precisely, for a first order transition one expects $\gamma=1$ and $\nu=1/3$
while for a second order transition $\gamma\neq 1$ \cite{Binder:1984}.

\begin{figure}[htb]
\includeEPSTEX{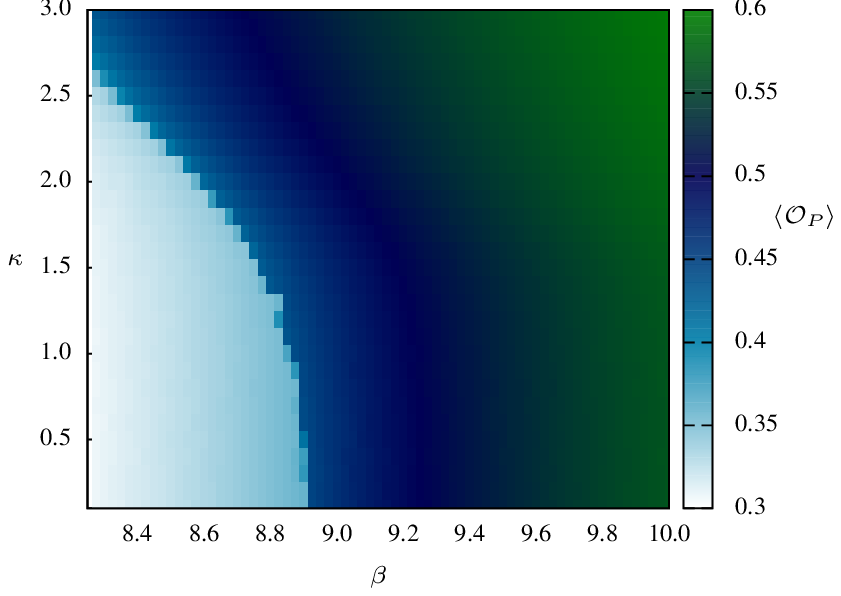}\hskip00mm
\includeEPSTEX{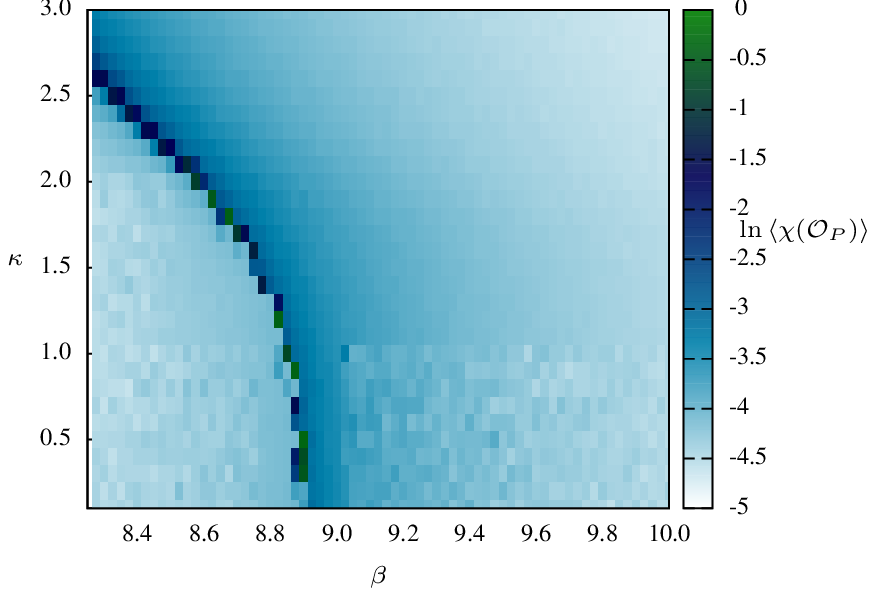}
\includeEPSTEX{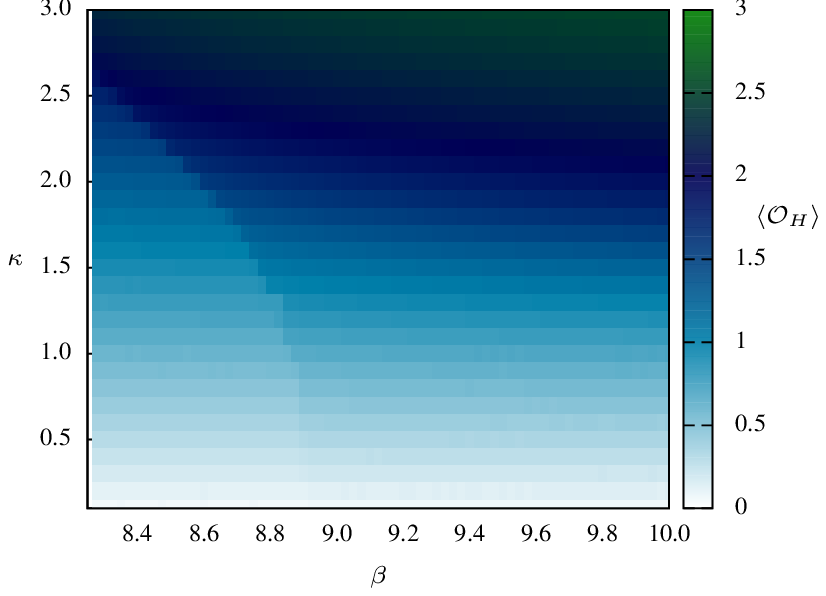}\hskip00mm
\includeEPSTEX{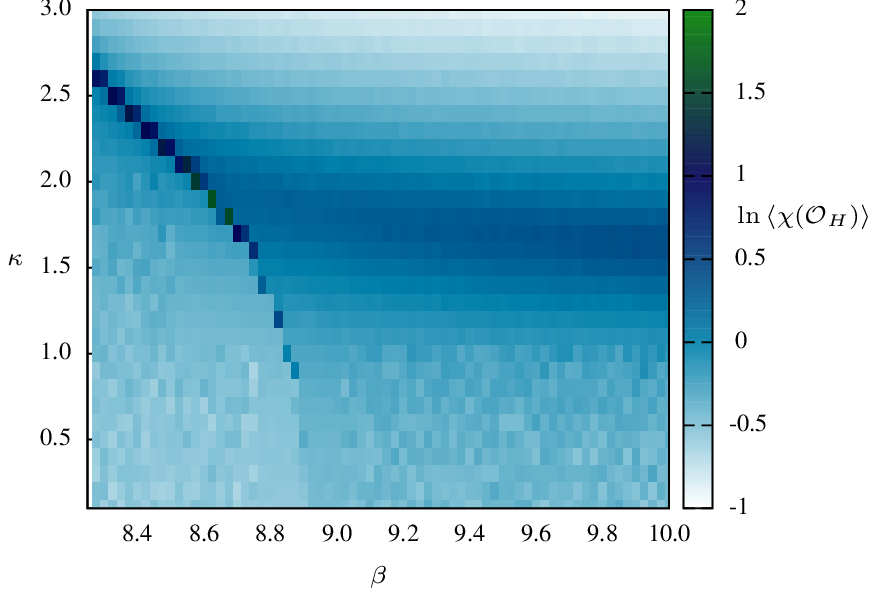}
\caption{Average plaquette, Higgs action and susceptibilities near the critical
point on $6^3\times2$ lattice.}
\label{fig:ympoldistkappa}
\end{figure}

The expectation values and logarithms of susceptibilities on
a small $6^3\times 2$-lattice are depicted in Fig.~\ref{fig:ympoldistkappa}.
The expectation value of a plaquette variable jumps at the deconfinement
transition line and the corresponding susceptibility is  peaked. 
This is in full agreement with the jump of the Polyakov loop across this transition line. 
The expectation value of the Higgs action and the corresponding susceptibility 
both spot the deconfinement transition well. But they also
discriminate between the $O(7)$ unbroken and broken phases. The data
on the small lattice point to a second order Higgs transition line in the YMH-model 
for all $\beta>\beta_{\rm deconf}(\kappa)$. This could imply that the second 
order line ends at the first order deconfinement  transition line. 
To determine the order of the Higgs transition line we consider
the finite size scaling of 
\begin{equation}
\label{eq:actionSuszeptibilities}
\chi(\gO_H)=\frac{\partial}{\partial \kappa} \left\langle \gO_H\right\rangle\quad
\hbox{and}\quad
\frac{\partial^2}{\partial^2 \kappa} \left\langle \gO_H\right\rangle
\end{equation}
for lattices up to $20^3\times 6$. The results presented below
show that the Higgs transitions are second order transitions.
Unfortunately we cannot exclude the possibility that the second order
line  turns into a crossover near the deconfinement transition line.

\begin{figure}[htb]
\includeEPSTEX{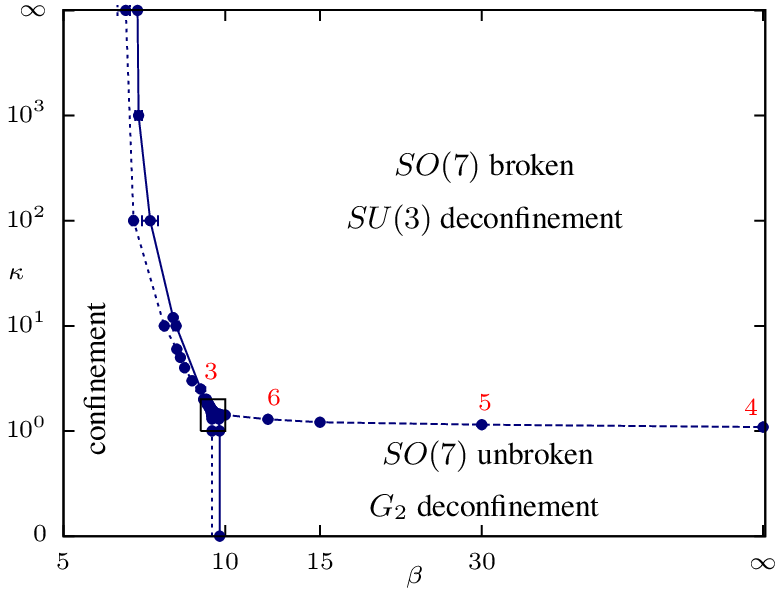}
\includeEPSTEX{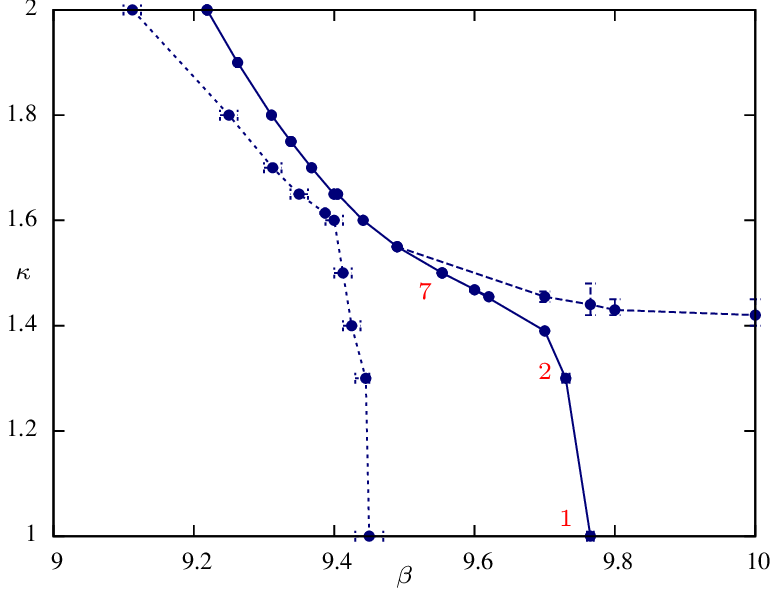}
\caption{Phase transition lines on a $16^3 \times 6$ lattice.
The solid line corresponds to the first order deconfinement
transitions, the dashed line to the second order Higgs transitions
and the dotted line to the left of the first order line
to the bulk transitions. The plot on the right shows
the details inside the small box in the plot
on the left where the transition lines almot meet.}
\label{fig:phasediagram16}
\end{figure}

Our results on the complete phase diagram in the $(\beta,\kappa)$-plane as 
calculated  on a larger $16^3\times 6$-lattice are summarized in 
Fig.~\ref{fig:phasediagram16}. We calculated histograms and susceptibilities
near the marked points on the transition lines in this figure. 
If the triple point exists then an extrapolation to the point 
where the confined phase meets both unconfined 
phases leads to the couplings $\beta_\text{trip}=9.62(1)$ and $\kappa_\text{trip}=1.455(5)$.
Near this  point the deconfinement transition is very weak, continuous
or absent  and thus we performed high-statistics simulations on larger lattices 
to investigate this region in parameter space more carefully. Some of our
results are presented in the following sections. Up to a rather small region
surrounding $(\beta_\text{trip},\kappa_\text{trip})$ we can
show that the deconfinement transition is first order and
the Higgs transition is second order. But we shall see that in a small region 
around this point the deconfinement transition is either second
order or absent.

\subsection*{The bulk transition}
The existence of a bulk transition in lattice gauge theories
at zero temperature can influence its finite temperature
behaviour. Such transitions are almost 
independent of the size of the lattice and are driven by 
lattice artifacts \cite{Halliday:1981te}.
Bulk transitions between the unphysical 
strong-coupling and the physical weak-coupling regimes
in lattice gauge theories is the rule 
rather than the exception. The strong coupling bulk phase 
contains vortices and  monopoles which disorder Wilson loops 
down to the ultraviolet length scale given by $a^2 \sigma \sim O(1)$ \cite{Lucini:2005vg,Brower:1981}. In the weak coupling phase 
the short distance physics is determined by aymptotic freedom
and $a^2 \sigma\ll 1$. Both $SU(2)$ and $SU(3)$ 
lattice theories exhibit a rapid crossover between the two phases 
which beomes more pronounced for $SU(4)$ \cite{Lucini:2005vg}. 
For $SU(N)$ with $N\geq 5$ the bulk transition is first order 
\cite{Lucini:2005vg}. $SU(3)$ lattice gauge theory with mixed
fundamental ($f$) and adjoint ($a$) actions shows a first order bulk transiton 
for large $\beta_a$ and small $\beta_f$. For decreasing $\beta_a$ the transition 
line terminates at a critical point and turns into a crossover touching the line
$\beta_a=0$. On lattices with $N_t=2$ the deconfinement transition line
joins the bulk transition line smoothly from below and for $N_t\ge4$ 
from above \cite{Caneschi:1981ik,Blum:1994xb}.
More relevant for us is the finding in  \cite{Cossu:2008} that 
the bulk transition in pure $G_2$ gauge theory at $\beta=9.45$ 
is a crossover \cite{Cossu:2008}.

\begin{figure}[htb]
\includeEPSTEX{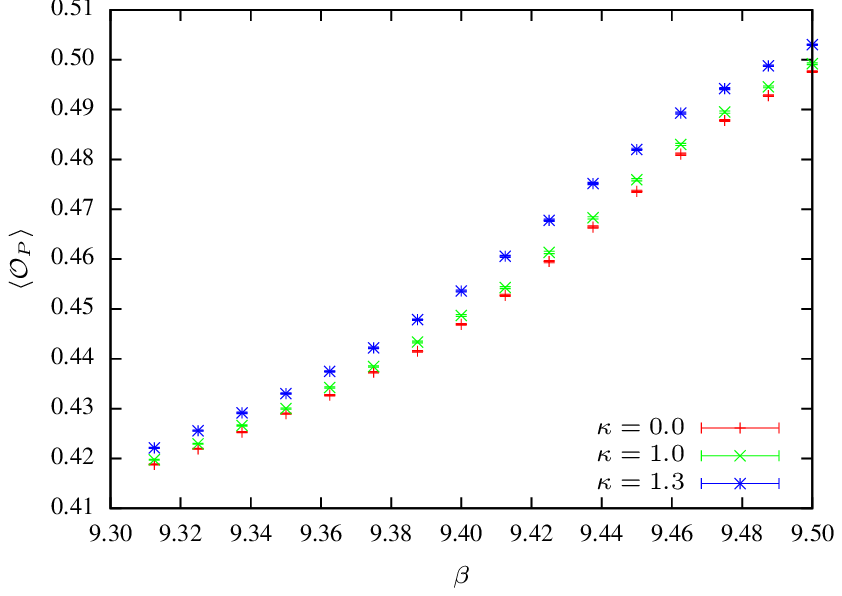}\hskip00mm
\includeEPSTEX{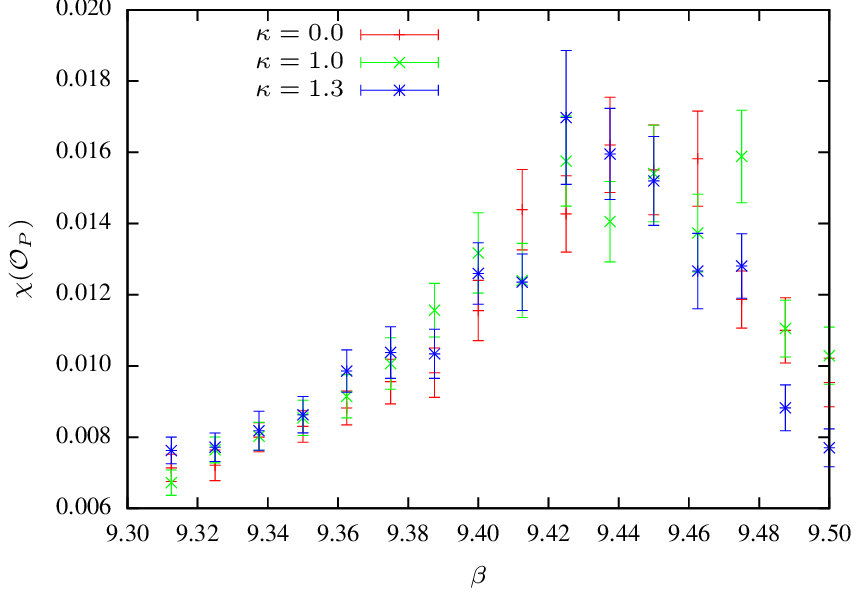}
\caption{Plaquette and susceptibility for small values of $\kappa$ near the
bulk transition on a $12^3 \times 6$ lattice.}
\label{fig:bulkSmall}
\end{figure}

\begin{figure}[htb]
\includeEPSTEX{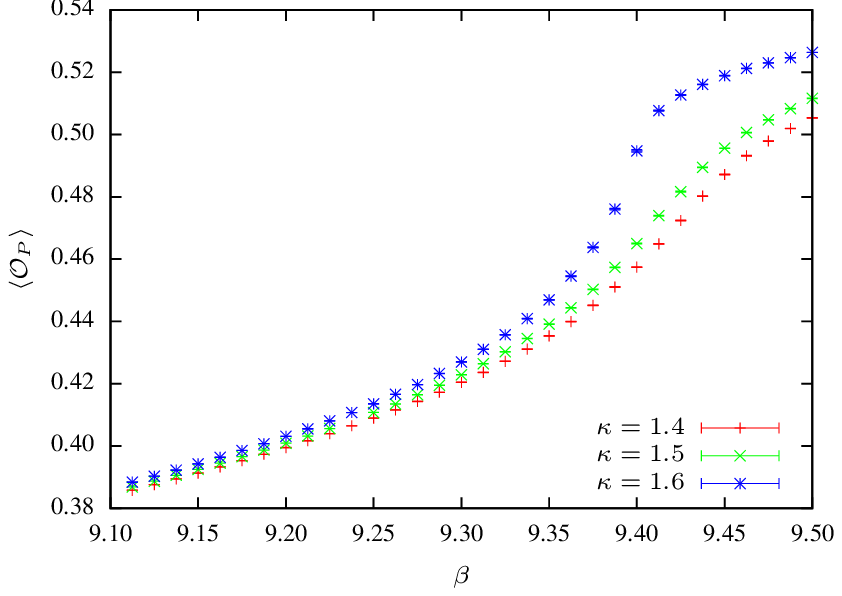}\hskip00mm
\includeEPSTEX{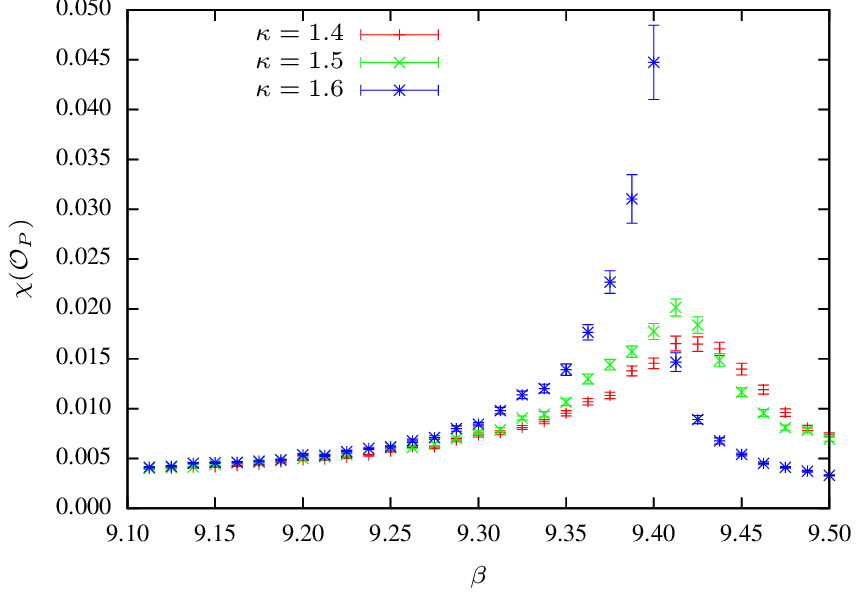}
\caption{Plaquette and susceptibility for intermediate values of $\kappa$ near
the bulk transition on a $12^3 \times 6$ lattice.}
\label{fig:bulkIntermediate}
\end{figure}

We have scanned the values for the plaquette variables and
Polyakov loops from the strong to the weak coupling regime to 
find a bulk transition that might interfere with the finite temperature
deconfinement  transition. For various values between $\kappa=0$ 
and $\kappa=\infty$ on a $12^3 \times 6$ and $16^3 \times 6$ 
lattice we determined the position and nature of the 
bulk transitions. In full agreement with \cite{Cossu:2008} we see a 
crossover at $(\beta,\kappa)\approx (9.44,0)$ which is visible as a broad peak
in the plaquette susceptibility depicted in the right panel of Fig.~\ref{fig:bulkSmall}. The Polyakov loop does not detect 
this crossover.
Note that for small $\kappa$ the position of the bulk transition does 
not depend on the hopping parameter which means that the bulk transition line
hits the line $\kappa=0$ vertically. Despite of the broad peak in
the susceptibility of the plaquette density are the bulk
and deconfinement transition cleary separated and
this agrees with the results in \cite{Bonati:2009pf}. 
In the region $1.3\leq\kappa\leq 1.6$ the critical coupling $\beta_c$ 
decreases with increasing $\kappa_c$ but the nature of the transition does 
not change much as can bee seen in Fig.~\ref{fig:bulkIntermediate}.
The plaquette density seems to be a continous function of $\beta$ 
and $\kappa$ and we conclude that the transition is still a crossover. 

\begin{figure}[htb]
\includeEPSTEX{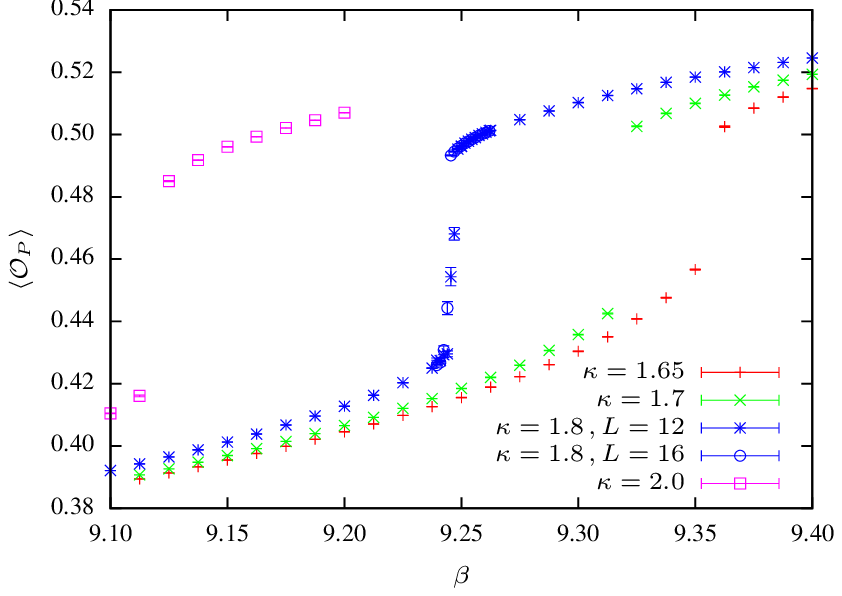}\hskip00mm
\includeEPSTEX{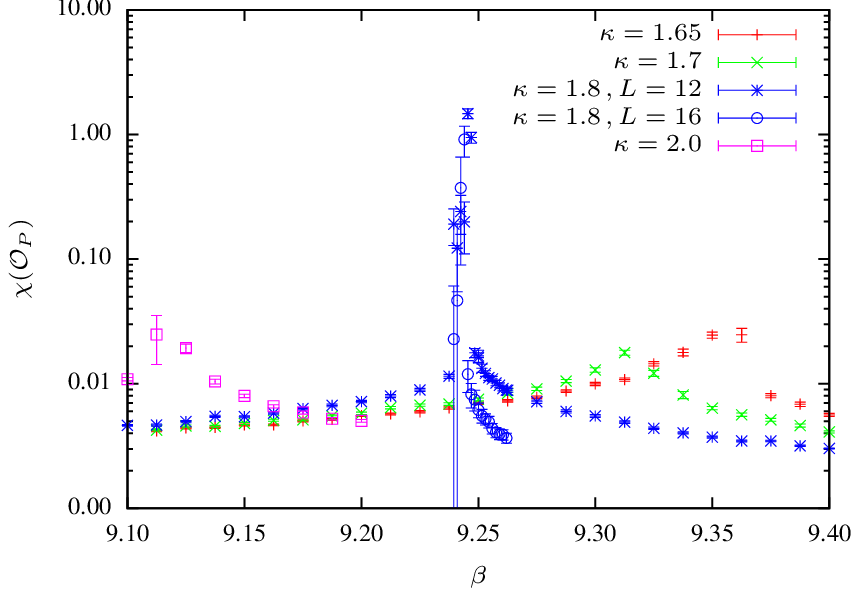}
\caption{Plaquette and susceptibility for intermediate values of $\kappa$ near
the bulk transition on $12^3 \times 6$ lattice.}
\label{fig:bulkFirst}
\end{figure}

Between $\kappa=1.6$ and $\kappa=1.65$ the peak in the bulk transition 
becomes pronounced. In this region the distance between the 
bulk and deconfinement transitions  becomes 
very small. Nevertheless we expect that the much localized bulk transition 
still does not interfere with the weak deconfinement transition. For values of
$\kappa$ between $1.65$ and approximately $2.5$ the position
of the bulk transition gets more sensitive to the hopping parameter and the distance to the 
deconfinement transition line increases again. The nature of the 
transition changes at the same time -- a large gap in the action density 
separates the strong  coupling from the weak coupling region. This is depicted in
Fig.~\ref{fig:bulkFirst}.
The many data points taken at $\kappa=1.8$ show that the size of the gap does 
not depend on the volume and this points to a first order  transition. 
The plots for the plaquettes and plaquette susceptibilites look very
much like the plots in Fig.~\ref{fig:bulkSmall}. For $\kappa
\gtrsim  2.5$ the situation changes again. The gap in the plaquette density closes
and the position of the bulk transition tends to  that of the bulk transition in $SU(3)$ gluodynamics 
which again is a crossover. 

\begin{figure}[htb]
\includeEPSTEX{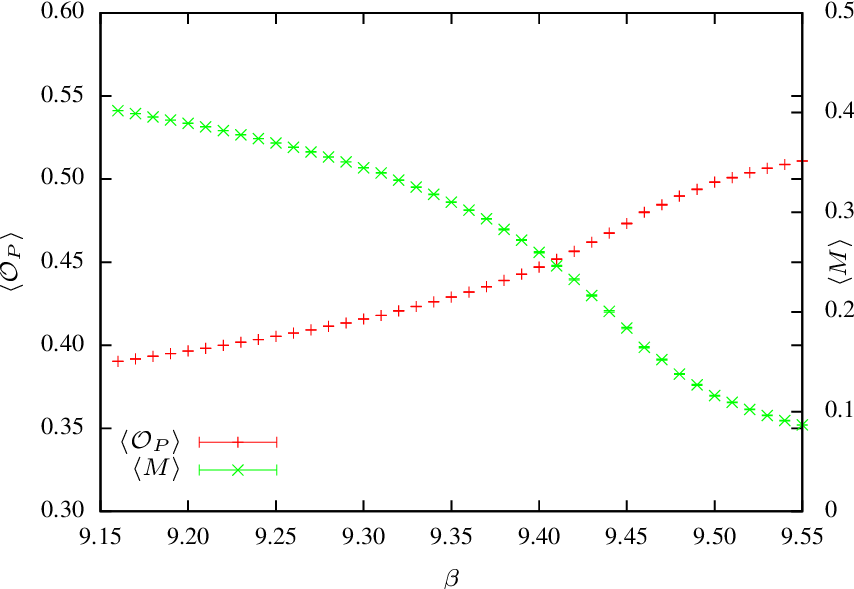}\hskip00mm
\includeEPSTEX{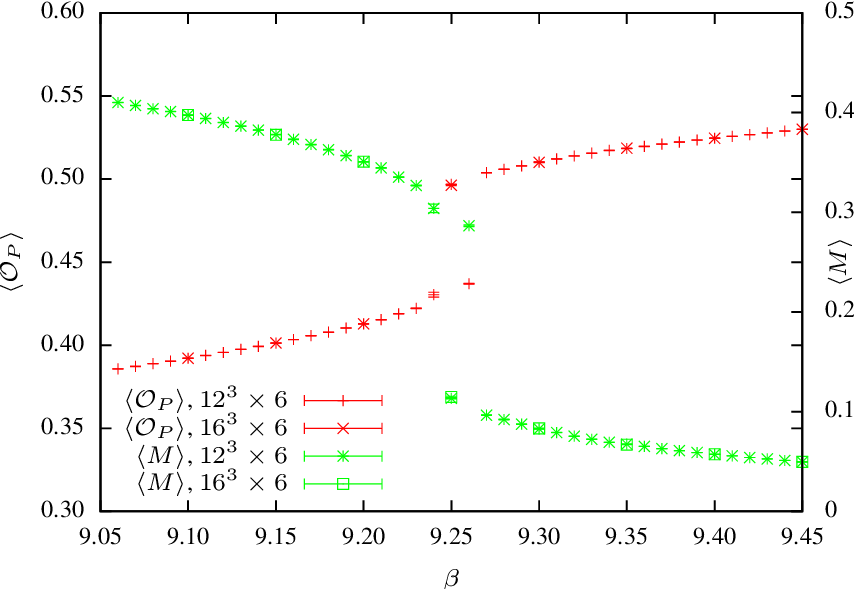}
\caption{Plaquette and Monopole density for $\kappa=0$ and $\kappa=1.8$ on a
$12^3 \times 6$ and $16^3 \times 6$ lattice.}
\label{fig:bulkMonopol}
\end{figure}

There is ample evidence that bulk transitions are driven by
monopoles on the lattice \cite{Halliday:1981te}. Thus
we calculated the density of monopoles \cite{Caneschi:1981ik}
as a function of $\beta$ for $\kappa=0$ and $\kappa=1.8$.
The density $M$ together with the plaquette variable
are plotted in Fig.~\ref{fig:bulkMonopol}.
For $\kappa=0$ they vary smoothly with $\beta$, as expected for a
cross-over, but for $\kappa=1.8$ they jump at the same 
$\beta\approx 9.25$. The height of the jump does not depend
on the lattice size, see Fig.~\ref{fig:bulkMonopol}, right panel.
Thus we find strong evidence that the bulk transition
is intimately related to the condensation of monopoles in the
strong coupling $G_2$ Higgs model.

Finally we would like to comment on the behaviour near $\kappa=1.6$.
Here the $G_2$ Higgs model behaves similar to $SU(3)$ gluodynamics
with mixed fundamental and adjoint actions.
The latter shows a first order bulk transition which turns 
into a crossover for small $\beta_a$. It seems that for  
$\kappa \gtrsim 1.6$ the massive $G_2$-gluons are heavy enough 
such that the approximate center symmetry of the unbroken $SU(3)$ 
is at work. This could explain why we find a first order 
transition for $\kappa \gtrsim 1.6$.

\section{The transition lines away from the triple point}
\label{sec:transitionLinesAway}
In this section we come back to the confinement-deconfinement
transition. Sufficiently far away from the suspected triple point at
$\beta_\text{trip}=9.62(1)$ and $\kappa_\text{trip}=1.455(5)$
the signals for first- and second order phase transitions are
unambiguous and are presented in this section. The measurements
taken near the would-be triple point are less conclusive and will be
presented and analysed in the following section.

\subsection*{The confinement-deconfinement transition line}
Already the histograms for the Polyakov loop show that the deconfinement 
transition is first order for values of the hopping parameter $\kappa$
in the intervals $[0,1.4]$ and $[1.7,\infty]$.
\begin{figure}[htb]
\includeEPSTEX{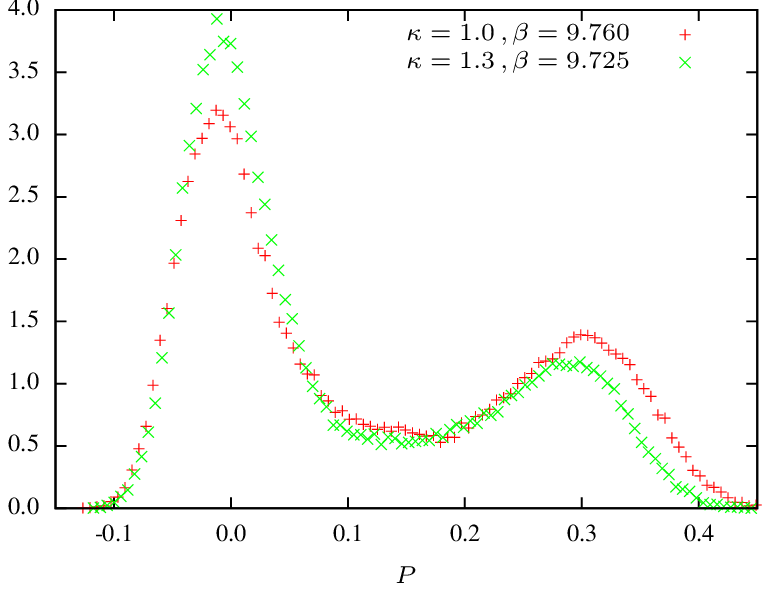}
\includeEPSTEX{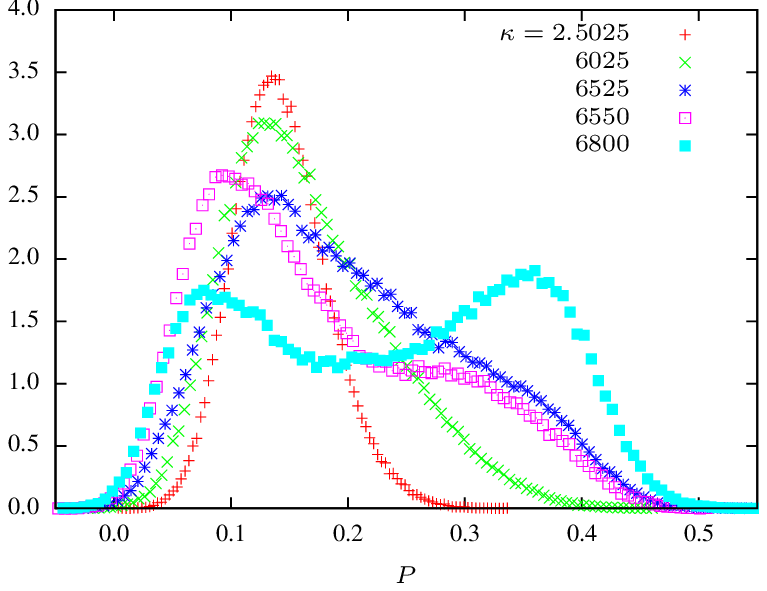}
\caption{Distributions of the Polyakov loop on a $16^3\times 6$-lattice.
Left panel: $(\beta,\kappa) = (9.76,1)$ and $(9.725,1.3)$; Right panel:
$\beta=9$ and various values of the hopping parameter.}
\label{fig:ympoldistkappa1}
\end{figure}
Two typical distributions for $\kappa=1.0$ and $\kappa=1.3$
corresponding to the points $1$ and $2$ in the phase diagram 
in Fig.~\ref{fig:phasediagram16}  are depicted in Fig.~\ref{fig:ympoldistkappa1}
(left panel). These and other histograms with $\kappa\lesssim 1.4$  show a
clear double peak structure near the transition line and are almost identical to the histogram for
$\kappa=0$. Similar results are obtained for larger hopping parameters
$\kappa\gtrsim 1.7$. 

In Fig.~\ref{fig:ympoldistkappa1} (Right panel) we plotted histograms of the
Polyakov loops for $\beta=9$ and hopping parameters in the
vicinity of $\kappa\approx 2.6$, corresponding to
point $3$ in Fig.~\ref{fig:phasediagram16}.
The histograms with $\kappa\leq 2.6525$ show peaks at almost
the same positions. The systems with these small values  of $\kappa$
are in the confined phase. For larger $\kappa$-values the peak moves
towards the 'would-be' center elements of the subgroup $SU(3)$ and
a second peak appears. Again the double-peak structure of
the distribution points to a first order transition. We varied the spatial 
sizes of the lattices and observed no finite size effects in the distributions 
for $N_s\geq 16$.
\subsection*{The Higgs transition line}
\begin{figure}[htb]
\includeEPSTEX{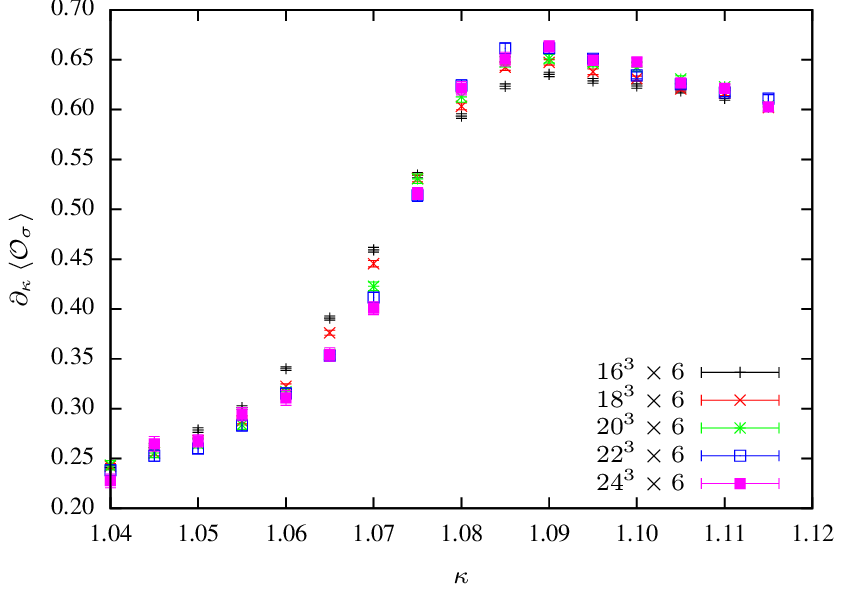}\hskip10mm
\includeEPSTEX{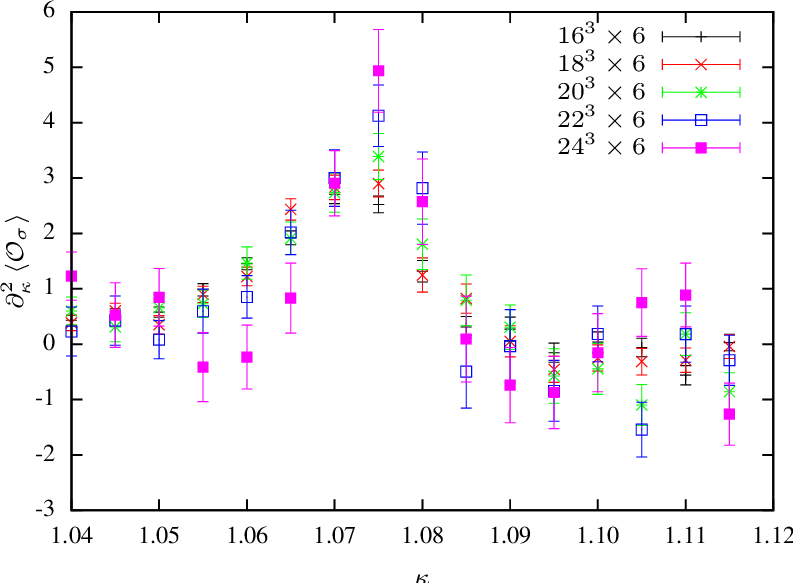}
\caption{The first and second derivative of the average sigma-model action 
for different spatial lattice sizes.}
\label{fig:so7sigma}
\end{figure}

For $\beta\to\infty$ the gauge degrees of freedom
are frozen and we are left with a nonlinear $O(7)$ sigma-model
which shows a second order transition from a $O(7)$-symmetric massive 
phase to a $O(6)$-symmetric massless phase. With the help of a cluster 
algorithm \cite{Wolff:1989} we updated the constrained scalar 
fields and calculated the susceptibility of
\begin{equation}
\gO_\sigma=\frac{1}{V}\sum_{x,\mu}\Phi_{x+\hat\mu}\Phi_x\label{goH}
\end{equation}
which is proportional to the sigma-model action $S_\sigma$ in (\ref{sigmaaction}),
\begin{equation}
 \chi(\gO_\sigma)=-\frac{1}{\kappa V}\,\partial_\kappa\langle S_\sigma\rangle.
\end{equation}
The results of our simulations on lattices with varying spatial sizes 
are depicted in Fig. \ref{fig:so7sigma}, left panel.
The susceptibility of the action becomes steeper as the spatial
volume increases while the peak of the (normalized) second derivative also
increases. This means that the system undergoes a second order transition
at $\kappa_c=1.075(5)$ (corresponding to point $4$ in
Fig.~\ref{fig:phasediagram16}) from a massive $O(7)$-symmetric phase with 
vanishing vacuum expectation value to a massless $O(6)$-symmetric 
phase with  non-vanishing expectation value. Actually the mean field theory 
for $O(n)$  models in $d$ dimensions predicts a second order transition at the 
critical coupling  $\kappa_{c,\rm mf}=n/2d$. For our model in $4$ dimensions
the mean-field prediction is $\kappa_{c,\rm mf}=7/8\approx 0.875$ and
is not far from our numerical value.

\begin{figure}[htb]
\includeEPSTEX{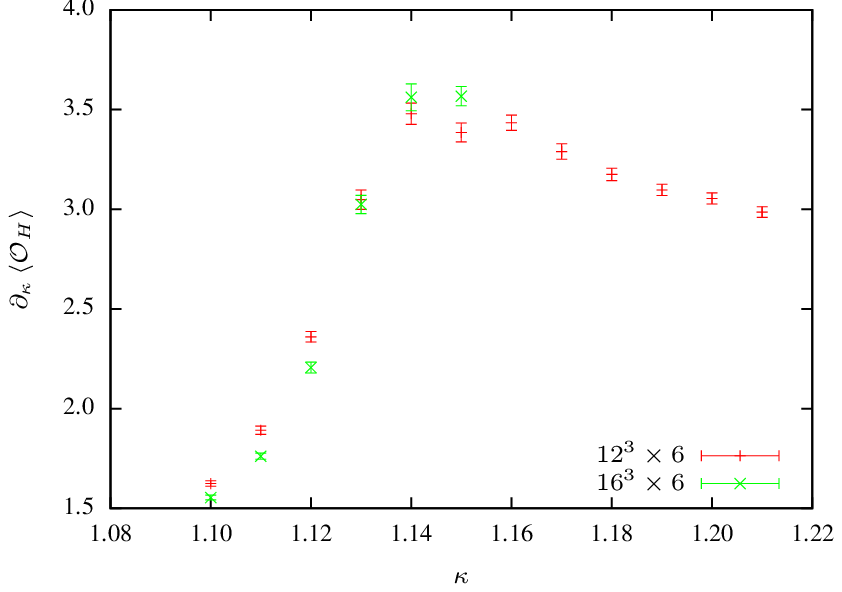}\hskip00mm
\includeEPSTEX{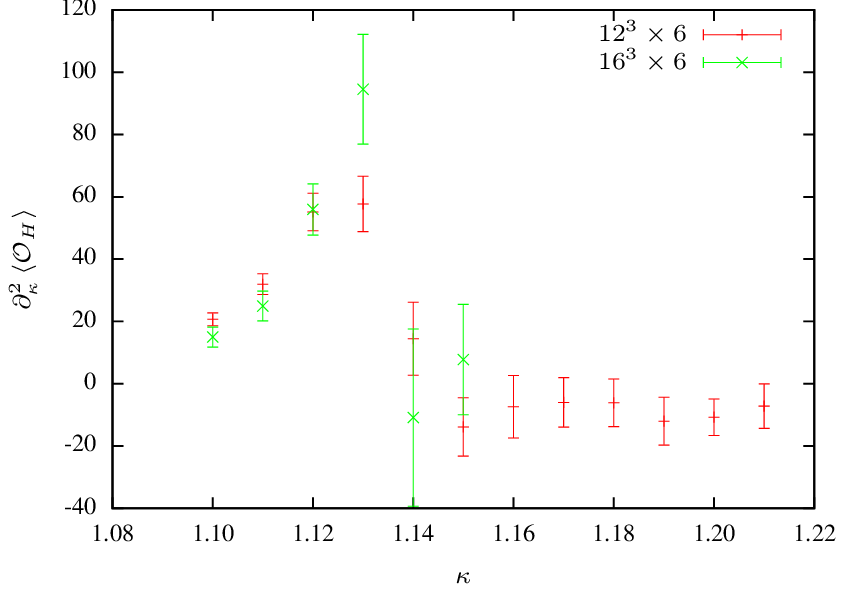}
\caption{First and second derivative of the average action with respect
to the hopping parameter for different spatial lattice sizes at $\beta=30$}
\label{fig:suszsigma1}
\end{figure}

\begin{figure}[htb]
\includeEPSTEX{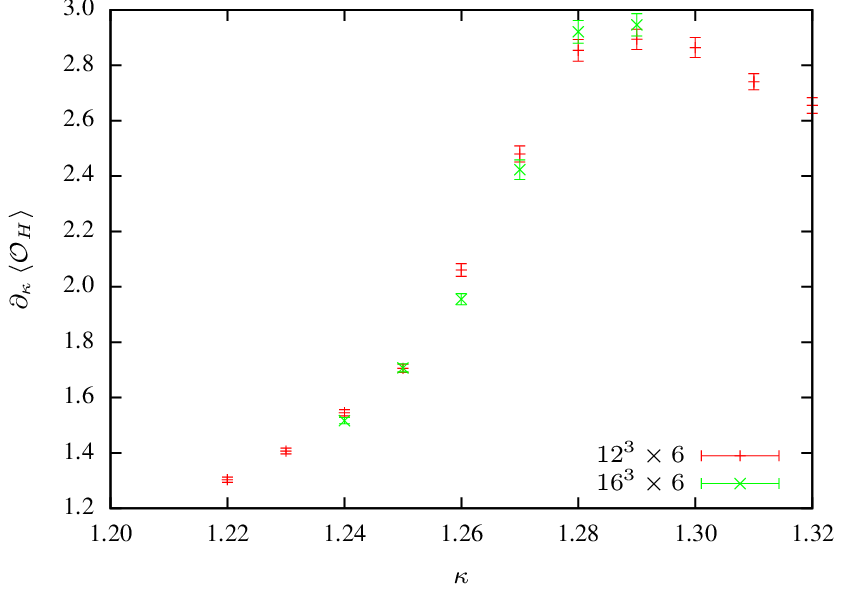}\hskip00mm
\includeEPSTEX{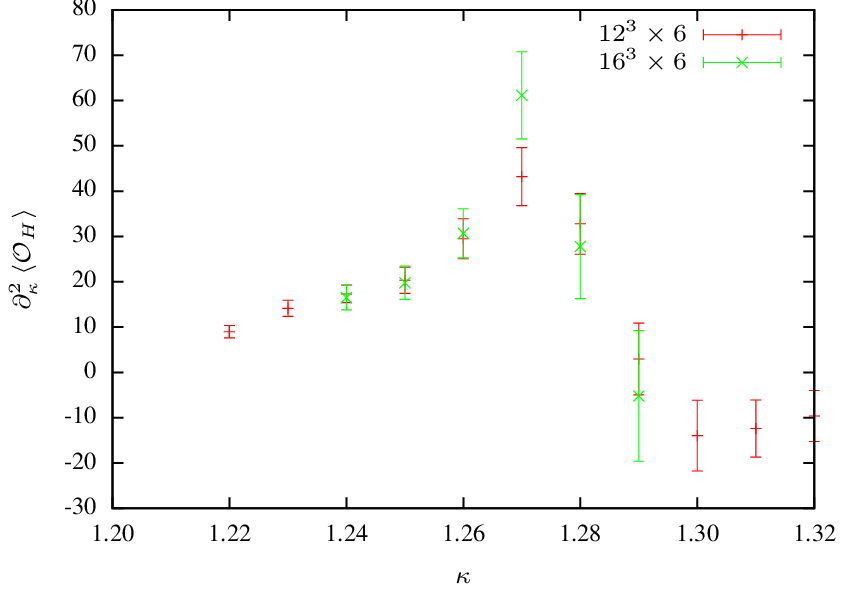}
\caption{First and second derivative of the average action with respect
to the hopping parameter for different spatial lattice sizes at $\beta=12$}
\label{fig:suszsigma2}
\end{figure}

For smaller values of $\beta$ the gauge degrees of freedom participate in 
the dynamics and $\partial_\kappa\langle S\rangle$ is now proportional to the susceptibility
of $\gO_H$ in  (\ref{operators}). The plots in Figs.~\ref{fig:suszsigma1} and \ref{fig:suszsigma2} 
show a similar behavior of the first and second
derivatives of the average Higgs action for $\beta=30$ and $12$, corresponding
to the points $5$ and $6$ in the phase diagram in Fig. \ref{fig:phasediagram16}.

Even for the smaller value $\beta=12$ we see that the susceptibility becomes 
steeper with increasing lattice size while the second derivative of the average action
increases. This already demonstrates that the second order transition at 
the aymptotic region $\beta\to \infty$ extends to smaller values of $\beta$.

\section{The transition lines near the triple point}
\label{sec:transitionLinesNear}
\begin{figure}[htb]
\scalebox{0.85}{\includeEPSTEX{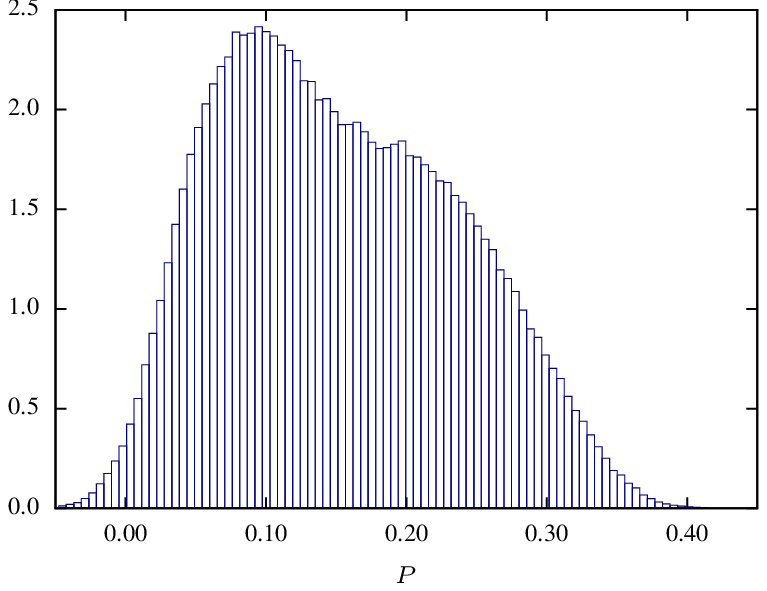}}\hskip00mm
\scalebox{0.85}{\includeEPSTEX{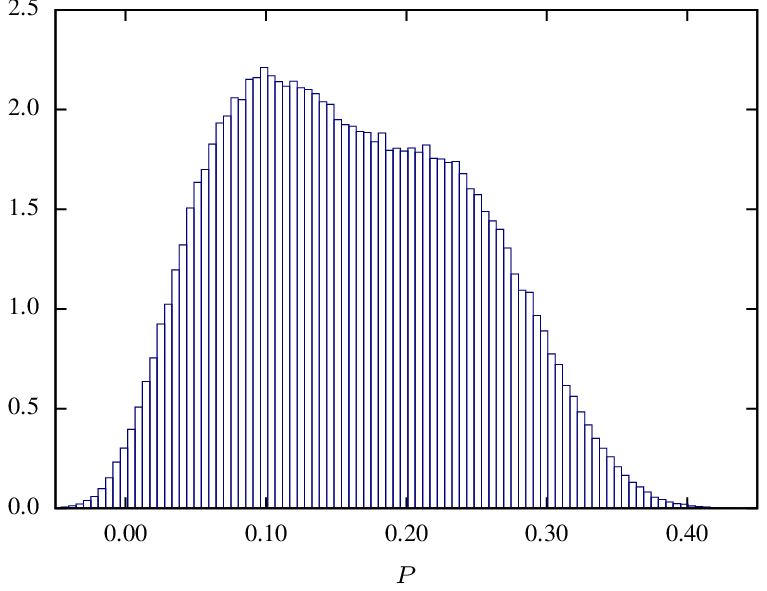}}
\scalebox{0.85}{\includeEPSTEX{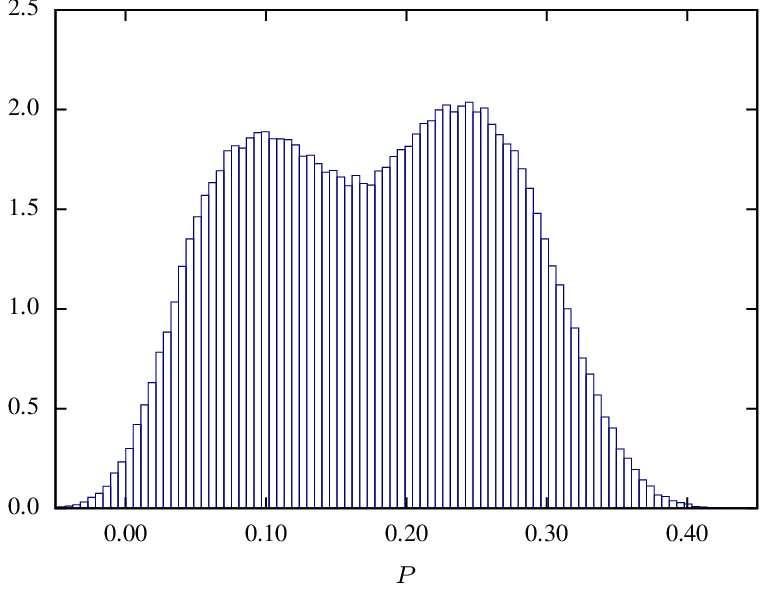}}\hskip00mm
\scalebox{0.85}{\includeEPSTEX{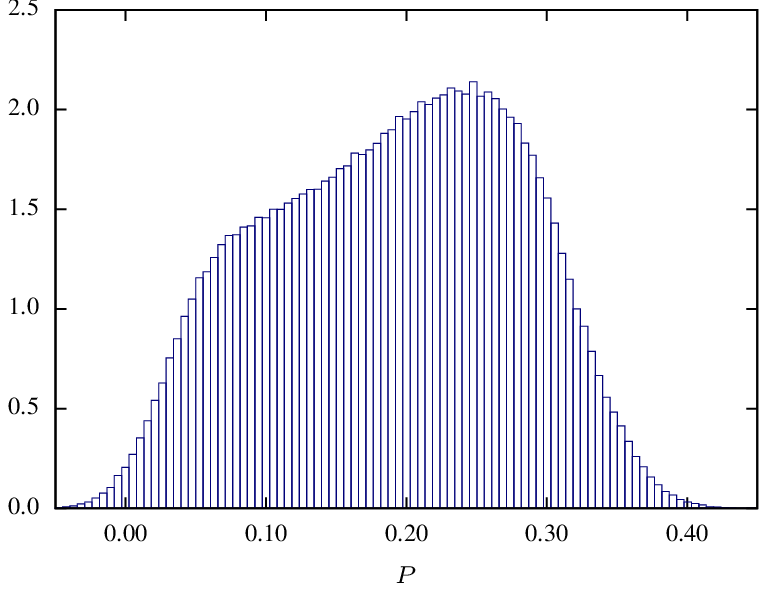}}
\caption{Distributions of the Polyakov loop at $\kappa=1.5$
where the transition is weakly first order 
on a $16^3\times 6$ lattice with $400\,000$
configurations for each histogram. Top left $\beta=9.5525$, top right $\beta=9.5535$,
bottom left $\beta=9.5540$ and bottom right $\beta=9.5550$ 
($\beta_c\approx 9.5535$).}
\label{fig:ympoldistkappa3}
\end{figure}

\begin{figure}[htb]
\scalebox{0.85}{\includeEPSTEX{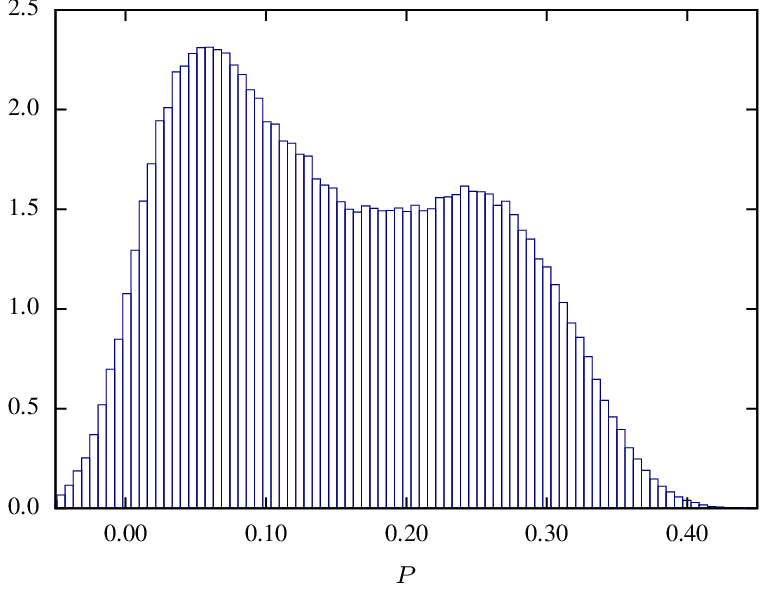}}\hskip00mm
\scalebox{0.85}{\includeEPSTEX{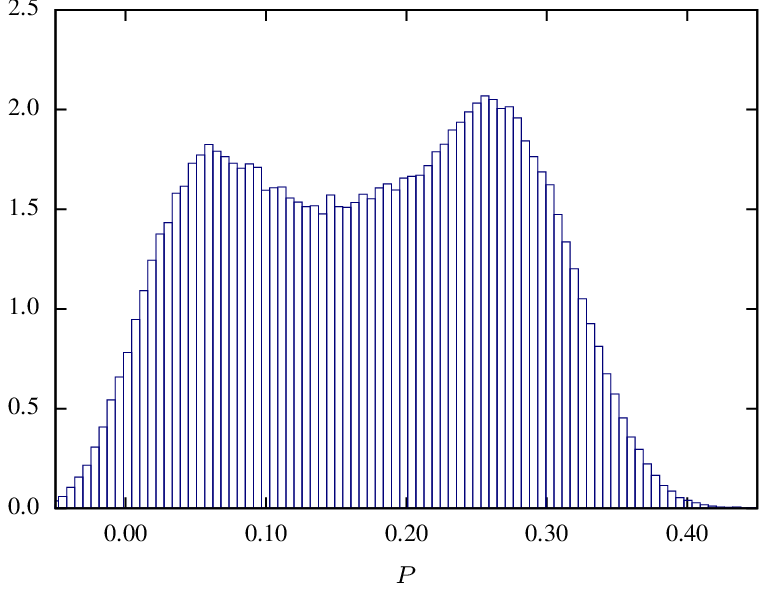}}
\scalebox{0.85}{\includeEPSTEX{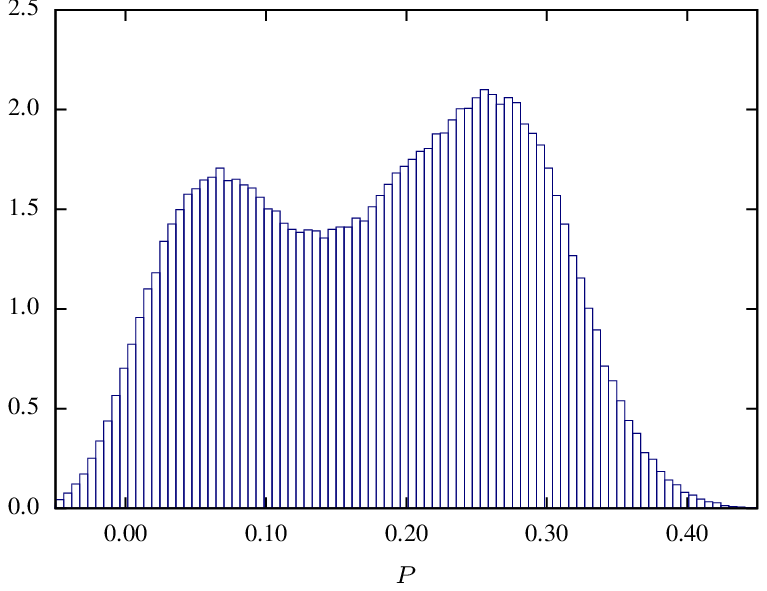}}\hskip00mm
\scalebox{0.85}{\includeEPSTEX{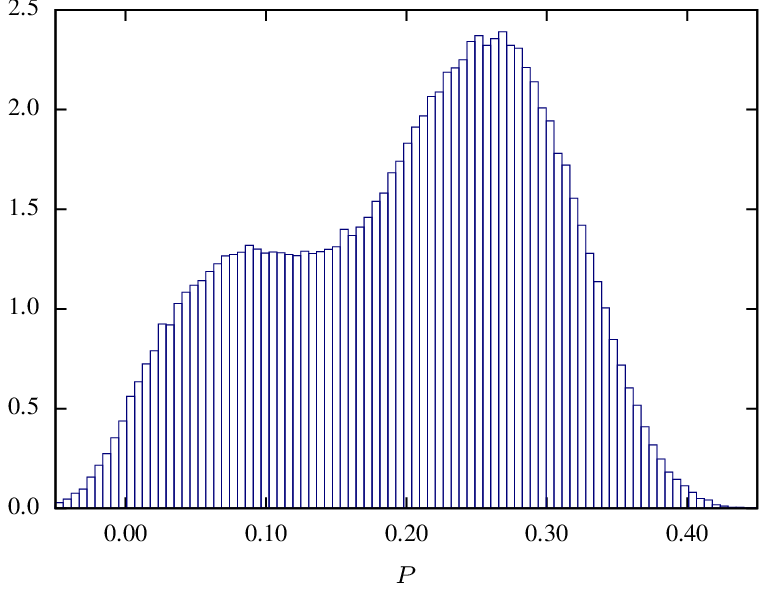}}
\caption{Distribution of the Polyakov loop at $(\beta,\kappa) =(9.6190,1.455) -
(9.6220,1.455)$ near the supposed triple point;
$400\,000$ configurations on $16^3\times 6$ lattice}
\label{fig:ympoldistkappa4}
\end{figure}

\begin{figure}[htb]
\scalebox{0.85}{\includeEPSTEX{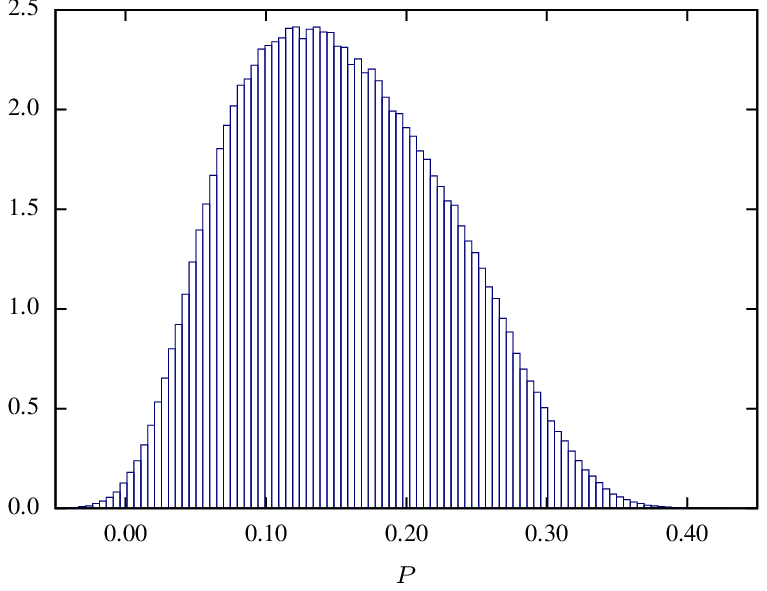}}\hskip00mm
\scalebox{0.85}{\includeEPSTEX{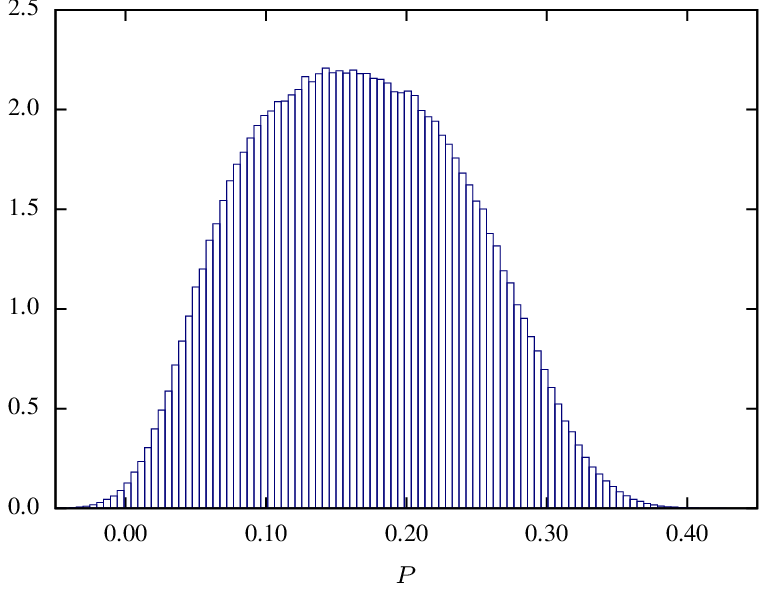}}
\scalebox{0.85}{\includeEPSTEX{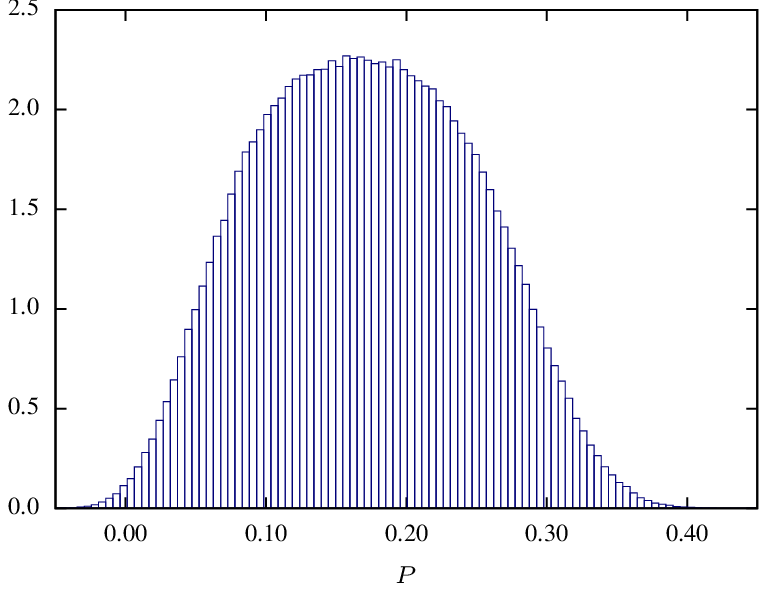}}\hskip00mm
\scalebox{0.85}{\includeEPSTEX{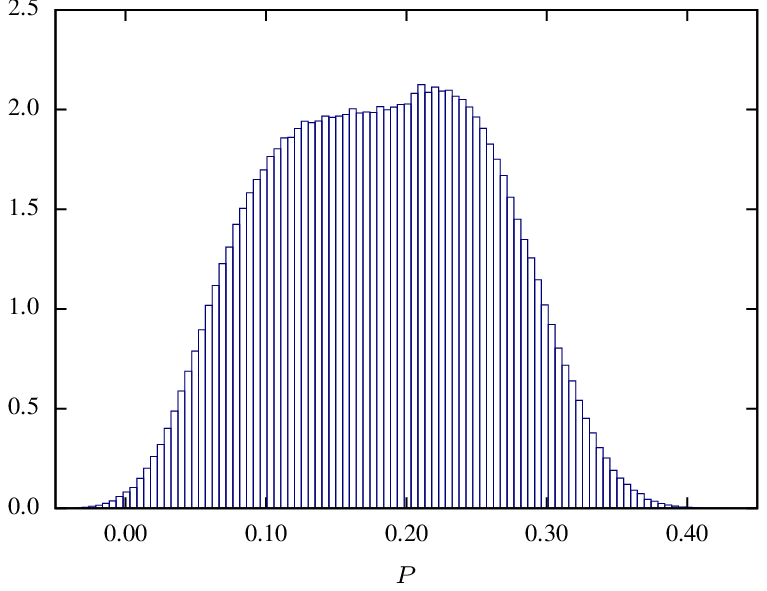}}
\caption{Distributions of the Polyakov loop at $\kappa=1.55$
where the transition is probably not first order 
on a $16^3\times 6$ lattice with $800\,000$
configurations for each histogram. Top left $\beta=9.4875$, top right
$\beta=9.4885$, bottom left $\beta=9.4895$ and bottom right $\beta=9.4905$ 
($\beta_c\approx 9.4885$).}
\label{fig:ympoldistkappa6}
\end{figure}

When the first order transition become weaker it becomes increasingly difficult
to distinguish it from a second order transition or a cross-over. For example, the
four histograms in Fig. \ref{fig:ympoldistkappa3} show distributions of the
Polyakov loop at point $7$ in the phase diagram depicted
in Fig.~\ref{fig:phasediagram16},
corresponding to $\kappa=1.5$ and $\beta$ varying between
$9.5525$ and $9.5550$. All histograms are computed from $400\,000$
configurations on a medium size $16^3\times 6$ lattice.
The histogram on top left shows a pronounced peak at $P\approx 0.1$,
corresponding to the value in the confined phase. With increasing
$\beta$ a second peak builds up at $P\approx 0.25$ corresponding
to a value in the unconfined phase. We have calculated more histograms
and conclude that the well-separated peaks in the distribution are of equal 
heights for $\beta_c\approx 9.5535$. At this point the Polyakov loop jumps from 
the smaller to the larger value. For even larger values of $\beta$ the 
second peak at larger $P$ takes over and the system is in the unconfined phase.
Although the histograms point to a weakly first order transition we can 
not rule out the possibility that the transition at $\kappa=1.5$ and 
$\beta\approx 9.5535$ is of second order. Later we shall see
that it is a first order transition. 
If we slightly decrease the value of $\kappa$, then the signal
for a first order transition is more pronounced. This is illustrated
in the Polyakov loop histograms depicted in Fig.~\ref{fig:ympoldistkappa4}.
If we again increase the value from $\kappa=1.5$ to
$\kappa=1.55$ the peak of the Polyakov loop does not jump at the transition
point at $\beta\approx 9.4885$. Instead it increases smoothly from
$P\approx0.12$ in the confinement phase to $P\approx 0.24$ in the deconfinement
phase, see Fig.~\ref{fig:ympoldistkappa6}. We conjecture that in this region 
of parameter space the first order transition turns into a continuous
transition or a cross-over which is later confirmed by an even more careful analysis.

We studied the size-dependence  of the average Polyakov loop, plaquette 
variable and Higgs-action per lattice site together with their 
susceptibilities. The following results are obtained on lattices with $N_t=6$ and 
spatial extends $N_s\in\{12,16,20,24\}$ and for $\beta=9.5535$. This
corresponds to points in the neighborhood of point $7$ in the phase 
diagram in Fig.~\ref{fig:phasediagram16}.

\begin{figure}[htb]
\includeEPSTEX{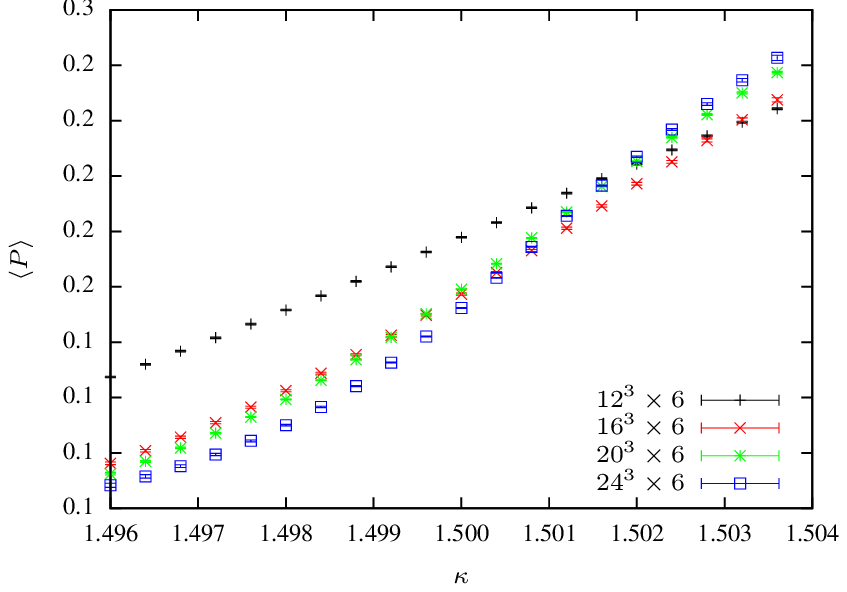}\hskip00mm
\includeEPSTEX{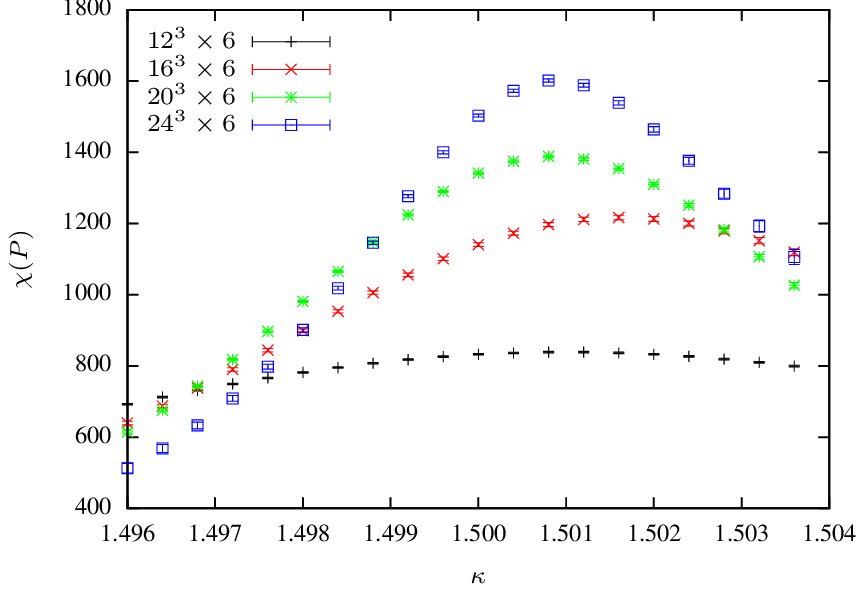}
\caption{Finite size scaling of Polyakov loop
and Polyakov loop susceptibility  at $\beta =
9.5535$}
\label{fig:ympoldistkappa5}
\end{figure}

Fig. \ref{fig:ympoldistkappa5} shows the $\kappa$-dependence of the Polyakov
loop and its susceptibility for the four different lattices. The measurements
have been taken at $20$ different values of the hopping parameter in
the vicinity of $\kappa=1.5$.
This way we cross the phase transition line vertically in the $\kappa$-direction
at the transition point $7$ in the phase diagram in Fig.~\ref{fig:phasediagram16}.
The $\kappa$-dependence has been calculated with
the reweighting method. Later we shall see that the peak of the susceptibility 
at $\kappa_c\approx 1.501$ scales  linearly with the volume. This linear dependence 
is characteristic for a first order transition.

\begin{figure}[htb]
\includeEPSTEX{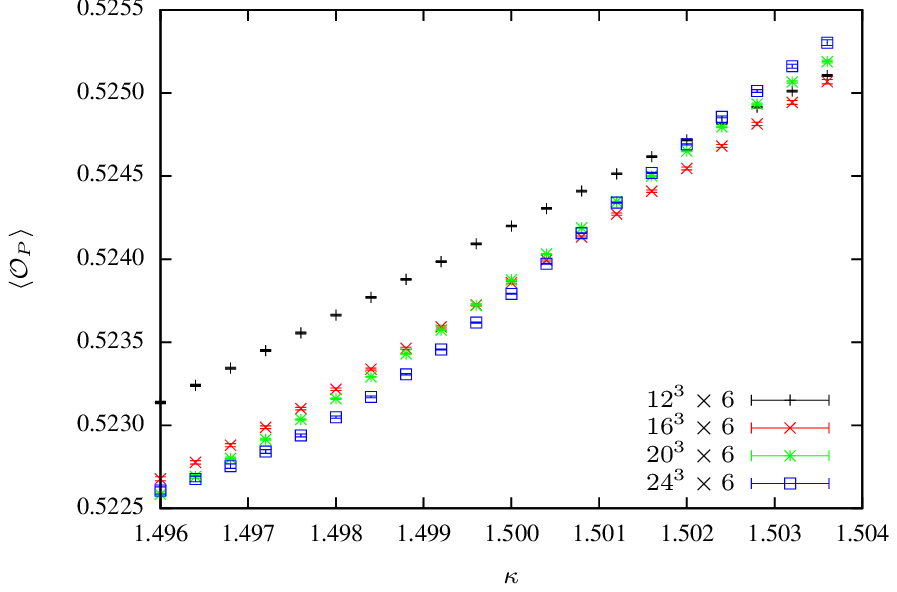}\hskip00mm
\includeEPSTEX{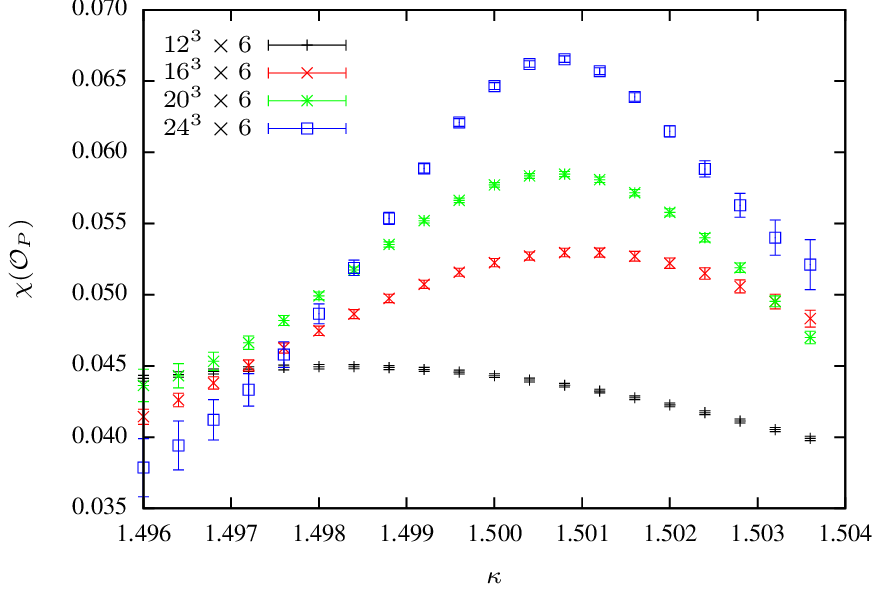}
\caption{Finite size scaling of the plaquette variable and
its susceptibility for $\beta = 9.5535$.}
\label{fig:ymactiondistkappa1}
\end{figure}

The plots in Fig.~\ref{fig:ymactiondistkappa1} show the 
$\kappa$-dependence of the average plaquette variable and the
corresponding susceptibility for  the four lattices. Again we observe that the 
susceptibility peak at $\kappa_c\approx 1.501$ increases linearly with the volume
of the lattice. Also note that on the small $12^3\times 6$ lattice the
peak in the susceptibility can hardly be seen.

\begin{figure}[htb]
\includeEPSTEX{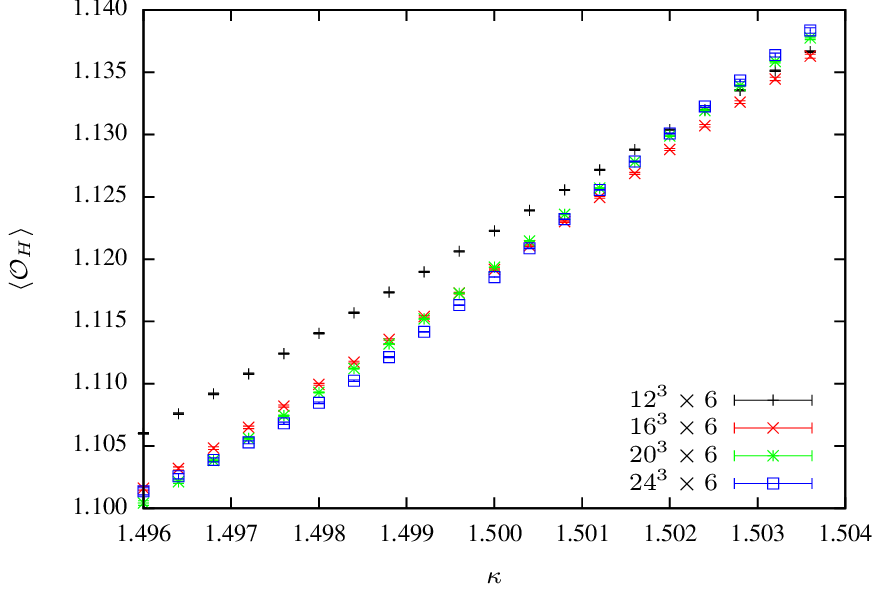}\hskip00mm
\includeEPSTEX{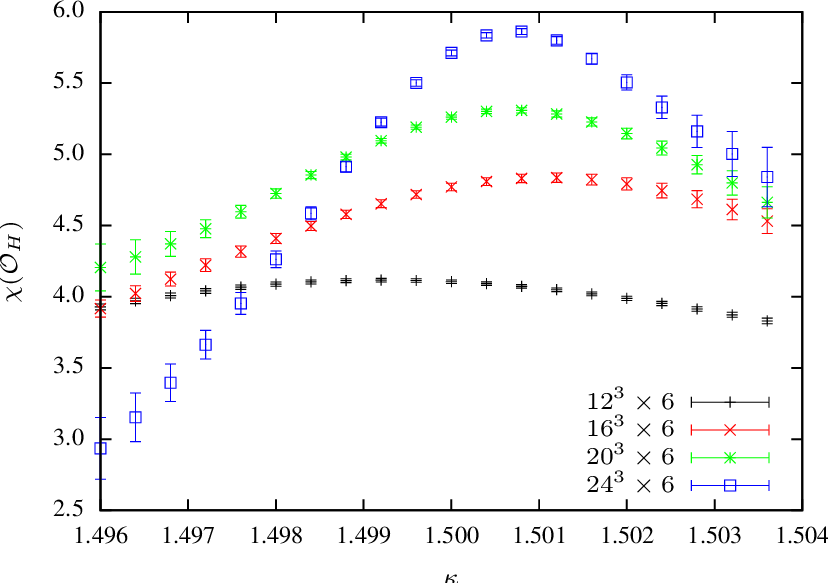}
\caption{Finite size scaling of Higgs action and its susceptibility for $\beta = 9.5535$.}
\label{fig:higgsdistkappa1}
\end{figure}

The two plots in Fig. \ref{fig:higgsdistkappa1} show the $\kappa$-dependence 
of the average Higgs action per lattice point and corresponding susceptibility.
Similarly as for the Polyakov loop and the plaquette we observe
a peak of the susceptibility at the same value $\kappa_c\approx 1.501$.

\begin{figure}[htb]
\includeEPSTEX{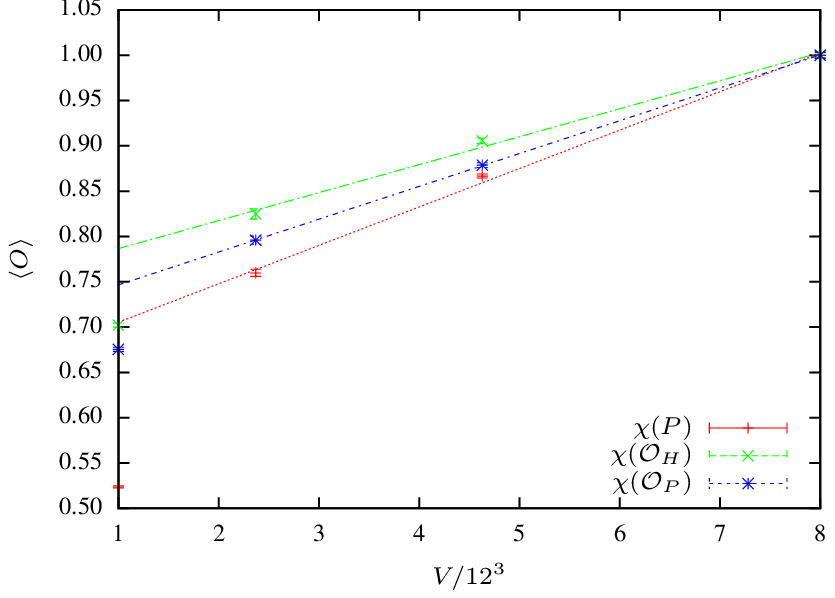}\hskip00mm
\includeEPSTEX{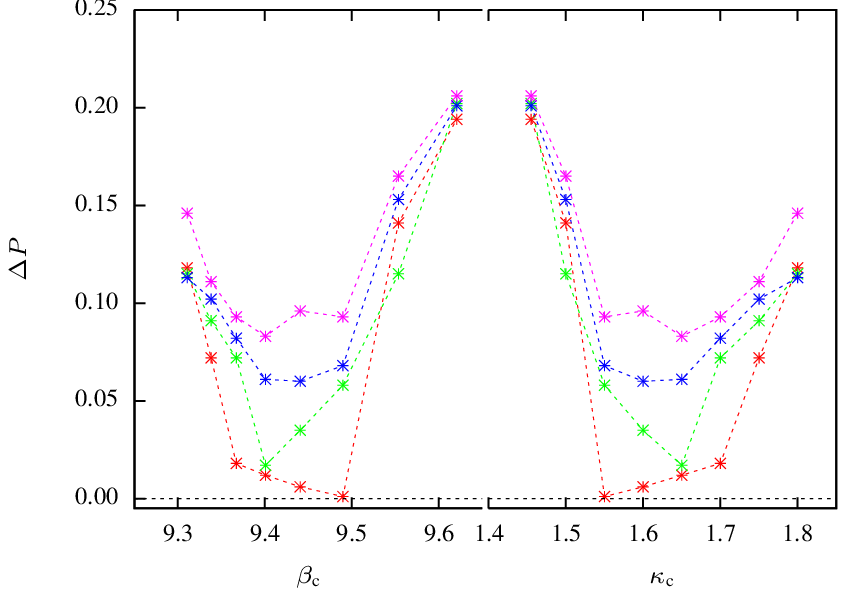}
\caption{Left panel: finite size scaling of the three susceptibilities at the transition
point with $\beta = 9.5535.$ The lines are fits to the peak values,
$\chi_{\rm max}(V)=a V+b$. Right panel: Difference of Polyakov loop in confined and unconfined phase at
the phase transition point for various couplings $\beta,\kappa$ and various
intervalls around the critical coupling coupling $\beta_c$, red: $\Delta
\beta=0.0005$, green: $\Delta \beta=0.0015$, blue: $\Delta \beta=0.0025$, pink:
$\Delta \beta=0.0035$, $\kappa$ is fixed ($\Delta \kappa=0$)}
\label{fig:chipeak_volume}
\end{figure}

To check for finite size scaling we investigated the susceptibilities
corresponding to the  Polyakov loop, plaquette variable and Higgs-action 
per site as a function of the volume. The results are plotted in Fig. \ref{fig:chipeak_volume},
left panel. For an easier comparison we normalized the data points by 
the peak value for the largest lattice with lattice size $N_s=24$.
The linear dependence of the peak susceptibilities on the volume is clearly visible 
for the larger three lattices and this linear dependence is predicted by
a first order transition \cite{Binder:1984}.
In recent studies of the lattice $SU(2)$ Higgs model in \cite{Bonati:2009pf}
it turned out that for $N_s=N_t\lesssim 18$ the maxima of the susceptibilities 
are well described by a function of the form $aL^4+b$, so that they seem
to scale linearly with volume, as expected for a first order transition at
zero temperature. Simulations on larger lattices revealed however, that 
the suceptibility peaks all saturate at larger values of $L$ and no singularities 
seems to develop in the thermodynamic limit. For the lattice $G_2$-Higgs model
considered in the present work we see no flattening of the peaks for 
larger lattices with $N_s$ up to $24$ and we interpret this as a signal for
a true first order transition.

Table \ref{tab:kritkappa}  shows the 
extrapolation of the critical hopping parameter to infinite volumes.
To that end we calculated for each lattice size the value $\kappa_c(V)$ 
at which the Polyakov loop-, plaquette- and Higgs action susceptibilities take 
their maxima. Note that on  the larger lattices with $N_s=20$ and $24$
the three critical hopping parameters are the same within statistical errors.
The infinite volume extrapolation yields the critical value $\kappa_c=1.5008$.

\begin{table}[htb]
\begin{tabular}{lcccc}
Volume & $12^3$  & $16^3$   & $20^3$  & $24^3$ \\ \hline
$\chi(P)$ & $1.5012$ & $1.5016$ & $1.5008$ & $1.5008$  \\
$\chi(\gO_H)$ & $1.4992$ & $1.5012$ & $1.5008$ & $1.5008$ \\
$\chi(\gO_P)$ & $1.4980$ & $1.5008$ & $1.5008$ & $1.5008$
\end{tabular}
\caption{Critical coupling $\kappa_c$ obtained from the maximum of
the susceptibility peaks of Polyakov loop, plaquette and Higgs action  for
different spatial volumes at $\beta=9.5535$}
\label{tab:kritkappa}
\end{table}

\subsection{The first order lines do not meet}
\begin{figure}[htb]
\includeEPSTEX{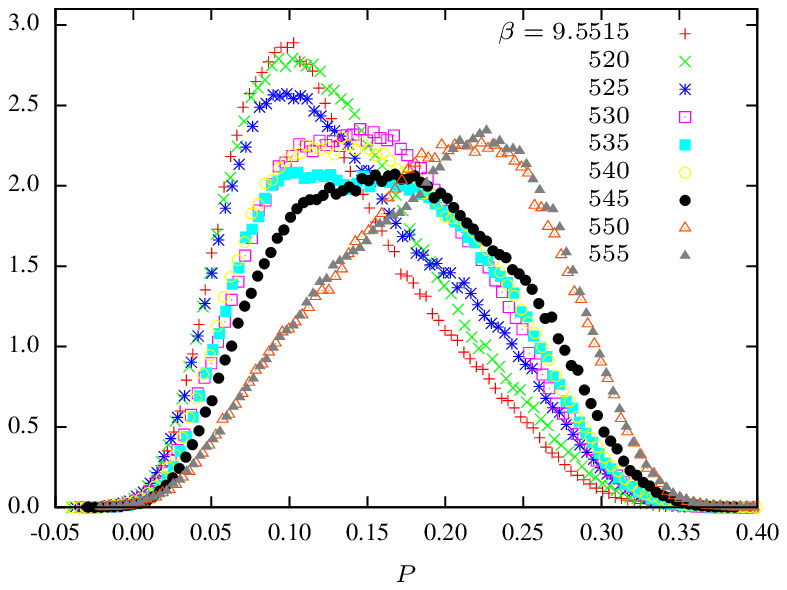}\hskip00mm
\includeEPSTEX{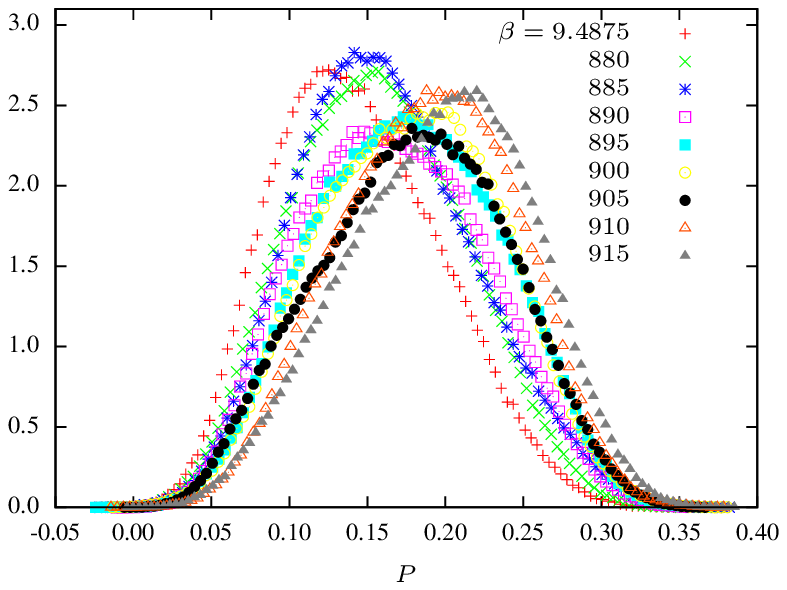}
\includeEPSTEX{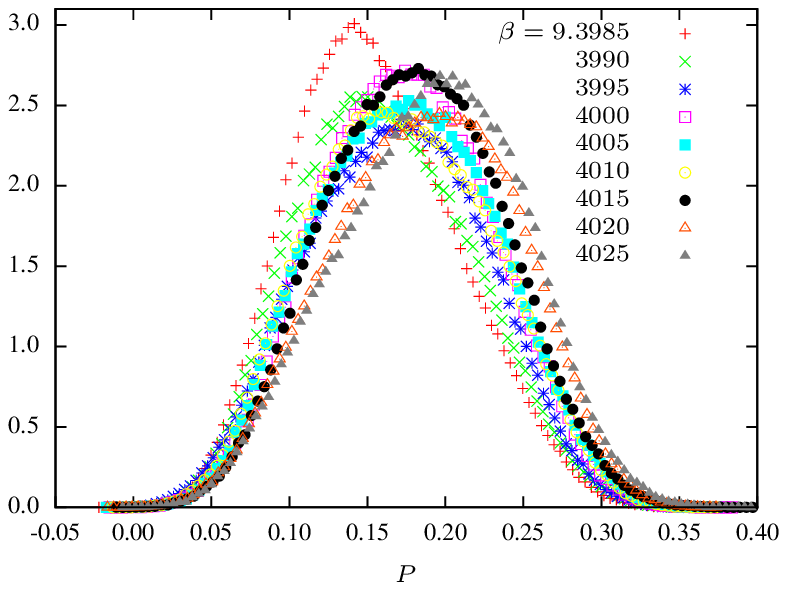}\hskip00mm
\includeEPSTEX{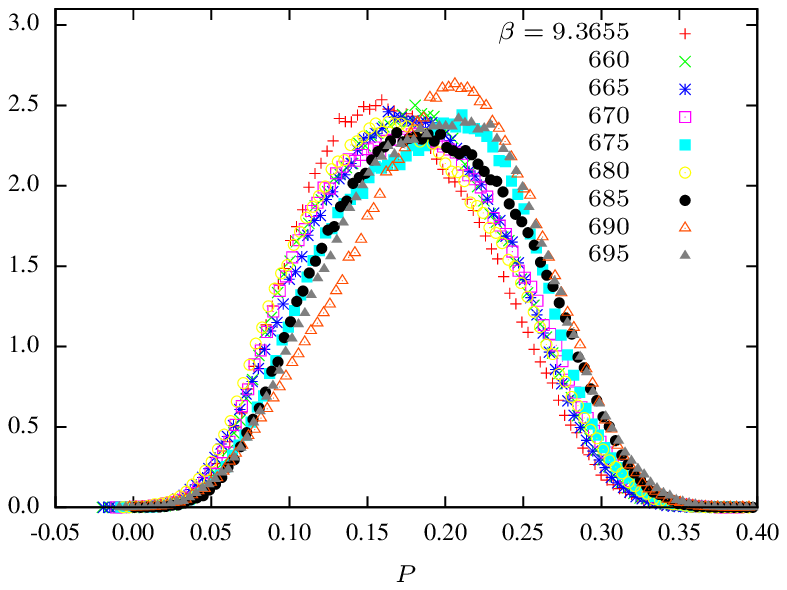}
\caption{Distribution of Polyakov loop near the phase transition point for
$\kappa=1.5$ (top left), $\kappa=1.55$ (top right), $\kappa=1.65$ (bottom left)
and $\kappa=1.7$ (bottom right) on a $20^3 \times 6$ lattice}
\label{fig:Histo_20}
\end{figure}

The previous results on the $16^3\times 6$ lattice
leave a small region in parameter space near  
$(\beta,\kappa)\approx (9.4,1.6)$, where the transition
may be continuous or where we can cross smoothly between 
the confined and unconfined phases. Since a jump of
the Polyakov loop expectation values in the infinite volume 
limit points to a first order transition we investigated the
quantity
\begin{equation}
\Delta P=\langle P\rangle_{\rm deconfined}-\langle P\rangle_{\rm confined}
\label{jump1}
\end{equation} 
more carefully. In the small parameter region we localized the
critical curve $(\beta_c,\kappa_c)$ with the histogram method. 
At the critical point is the height of the confinement peak
equal to the height of the deconfinement peak. For fixed
$\kappa_c$ we crossed the transition line by increasing the inverse gauge
coupling. Then we measured the maximal jump as a function of the step size
$\Delta\beta$ for one step size below and one above $\beta_c$. For
a first order transition the jump should not depend much on
$\Delta\beta$ whereas for a continuous transition or a 
cross-over $\Delta P$ should decrease with decreasing 
$\Delta\beta$. The results on a $16^3\times 6$ lattic are 
depicted in Fig.~\ref{fig:chipeak_volume} (right
panel). We see that for $9.35\lesssim \beta_c\lesssim 9.52$ corresponding 
to $1.52\lesssim\kappa_c\lesssim 1.72$ the jump approaches zero 
with shrinking step size and this clearly points to second order 
confinement-deconfinement  transitions or cross-overs in these 
small parameter regions. Simulations on a larger $20^3 \times 6$ lattice
confirm these results. Fig.~\ref{fig:Histo_20} shows histograms of
the Polyakov loop for $\kappa$-values between $1.5$ and $1.7$.
At $\kappa=1.5$ we still observe a weakly first order transition which 
turns into a continuous transition or crossover for $1.5<\kappa\leq 1.7$.
Within the given resolution in parameter space the window
is the same as on the $16^ 3\times 6$ lattice. Since the critical
couplings for spatial volumes beyond $20^3$ do not change we 
conclude that the gap will not close in the infinite volume limit.
This shows that the two first order lines emanating from $\kappa=0$
and $\kappa=\infty$ do not meet.

Here the question arises whether such a gap in the first order line 
between the confined and unconfined
phases is expected. The celebrated
Fradkin-Shenker-Osterwalder-Seiler theorem  
\cite{Osterwalder:1977pc,Fradkin1979}, originally
proven for the $SU(N)$ Higgs-model with scalars in the fundamental
representation, says that there is no complete separation between 
the Higgs- and the confinement  regions. Any point deep 
in the confinement regime and any point deep in the Higgs regime
are related by a path such that Green's functions of local, gauge
invariant operators vary analytically along the path. Thus there is 
no abrupt change from a colorless to a color-charged  spectrum.
This is consistent with the fact that there are only color singlet 
asymptotic states in both 'phases'.

The proof of the theorem relies crucially on using a completely-fixed 
unitary gauge. A complete gauge fixing is not possible with scalars
in the adjoint representation of $SU(N)$ since these scalars 
are center blind.
Thus the theorem does not hold for adjoint scalars and indeed, with  adjoint 
scalars there exits a phase boundary separating the Higgs and confined
phases. It is not completely obvious what these results tell us about 
the phase diagram of the $G_2$ Higgs model. The center of $G_2$ is trivial and the
$14$-dimensional adjoint representation is just one of the two 
fundamental representations. Since there is no need to break
the center one may conclude that the confinement-like regime
and the Higgs-like regimes are analytically connected. 
In addition, for large values of the hopping parameter the center of the 
corresponding $SU(3)$ gauge theory is explicitly broken by the scalar 
fields, simililarly as for the $SU(3)$ Higgs model with scalars
in the fundamental representation. These arguments suggest
that there exist a smooth cross-over between the confining
and Higgs phases. But one important assumption of the
Fradkin-Shenker theorem is not fulfilled for the $G_2$ Higgs model.
The theorem assumes that there exists no transition for large $\kappa$. 
Then at large $\kappa$ one can move 
from large to small $\beta$ and then at small $\beta$ further on to 
small values of $\kappa$ without hitting a phase transition.
Clearly this is not possible for the $G_2$ Higgs model such that 
not all assumption of the theorem hold true.

\FloatBarrier

\section{Conclusions}
\label{sect:conclusions}
\begin{figure}[htb]
\scalebox{1.2}{\includeEPSTEX{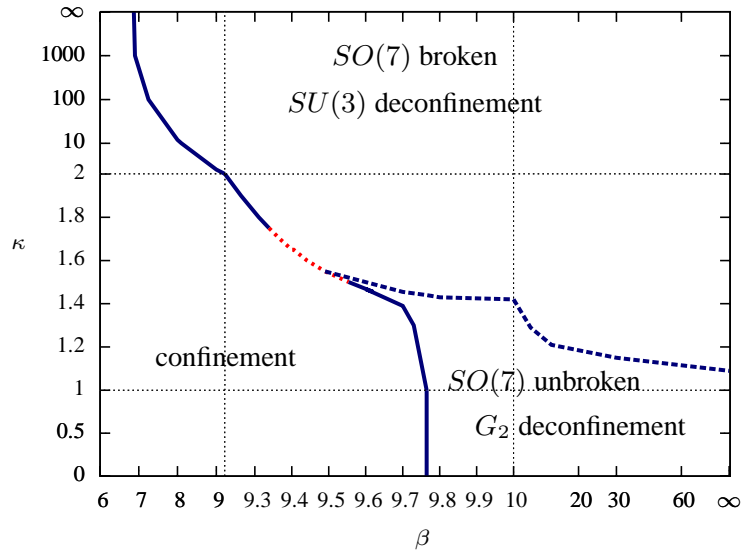}}
\caption{Complete phase diagram in the $(\beta,\kappa)$-plane
on a $16^3 \times 6$ lattice.
The neighbourhood of the 'would-be triple point' is very much enlarged
and the variable scale in the diagram is responsible for the cusps in the transition
lines. The solid line indicates a first order
transition, the dashed line (blue) a second
order transition and the dotted line (red) a
second oder transition or a crossover.}\label{fig:phasediagram16_summary}
\end{figure}

\noindent
With a new and fast LHMC-implementation for the exceptional
$G_2$ Higgs model we calculated the full phase
diagram in the coupling constant plane spanned by the hopping
parameter $\kappa$ and inverse gauge coupling $\beta$.
First we confirmed the proposed and earlier seen 
\cite{Pepe:2006er,Cossu:2008}  first order transition for pure $G_2$-gluodynamics 
which corresponds to the line $\kappa=0$ in the phase diagram of the Higgs model. 
A first analysis on smaller lattices indicated that this
first order transition is connected to the first order
deconfinement transition in $SU(3)$-gluodynamics, corresponding
to the limit $\kappa\to\infty$, by a smooth curve of first
order transitions. The same analysis spotted another curve of 
second-order transitions emanating from $\beta\to\infty$ and
meeting the first order line at a triple point. For this first
analysis we calculated histograms for the Polyakov
loop, Higgs-action and plaquette action. To identify the
second order transition line we studied the finite size
scaling of various susceptibilities and the second derivative
of the action with respect to the hopping parameter.
The final result of our analysis on a $16^3\times6$ lattice is depicted in 
Fig.~\ref{fig:phasediagram16_summary}. Note that 
the tiny region in the vicinity of the would-be triple 
point is very much enlarged in this figure.
In this tiny region in the $(\beta,\kappa)$-plane where the order 
of the transition could not be decided we studied the slope
of $\langle P\rangle$ in the vicinity of the suspected transition.
The simulations show that the two first-order curves emanating
from the lines with $\kappa=0$ and $\kappa=\infty$ end before they 
meet. The two curves could be connected by a line of second-order
transition or they could end at two (critical) endpoints in
which case the confined and unconfined phases are smoothly
connected. If indeed there exists a cross-over in $G_2$ Higgs
model at a finite value of the hopping parameter then the gauge
model behaves very similar to QCD with massive quarks.

To finally answer the question about the behavior of $G_2$ 
Higgs model theory in the vicinity of the 'would-be triple point' 
at $(\beta,\kappa)\approx (9.4,1.6)$ further simulations with
an even higher statistics and a more sophisticated analysis 
of the action susceptibilities may be necessary.   Since we already used an
efficient (and parallelized) LHMC-algorithm and much CPU-time to arrive 
at the results presented in the work this will not be an easy task.
Earlier studies of the susceptibility peaks in the simpler $SU(2)$-Higgs 
model on smaller lattices pointed to a first order transition at 
$\beta\lesssim 2.5$. Recent simulations  on larger lattices in \cite{Bonati:2009pf} 
showed that the susceptibility peaks do not scale with the volume
such that there is actually no first order transition for these small 
values of $\beta$.  We have seen no flattening of the peaks with the 
increasing volumes for $N_s\leq 24$ and conclude that the solid line
in Fig.~\ref{fig:phasediagram16_summary} is a first order line.
But of course we cannot exclude the possibility that the correlation
length is larger as expected and that simulations on even larger
lattices are necessary to finally settle the question about
the position and size of the window connecting the confined with the
unconfined phase. This will not be easy and thus it would be 
very helpful to actually prove that the confining and Higgs phases 
of $G_2$ can be connected analytically, perhaps with similar arguments 
as they apply to $SU(N)$ Higgs models with matter in the fundamental 
representations \cite{Osterwalder:1977pc,Fradkin1979}.

\begin{acknowledgments}
\noindent
We thank Philippe de Forcrand, Christof Gattringer,
Kurt Langfeld, \v{S}tefan Olejn\'{i}k, Uwe-Jens Wiese and, expecially, 
Axel Maas for interesting discussions or useful comments. This work has 
been supported by the DFG-Research Training Group ''Quantum- and 
Gravitational Fields'' GRK 1523 and the DFG grant Wi777/10-1.
The simulations in this paper were carried out at the Omega-Cluster 
of the TPI.
\end{acknowledgments}

\renewcommand{\eprint}[1]{ \href{http://arxiv.org/abs/#1}{[arXiv:#1]}}

\end{document}